\documentclass[12pt]{iopart}
\usepackage[T1]{fontenc}

\usepackage{url}

\usepackage{hyperref}
\usepackage{soul}
\usepackage{caption}
\usepackage{subcaption}
\usepackage{amssymb}
\usepackage{graphics,graphicx,color}
\usepackage{dcolumn,bm}
\begin{document}

%\title[Interplay between landscape structure and dynamics in the characterization of DNNs]{Interplay between landscape structure and dynamics in the characterization of Deep Neural Networks}

\title[Subaging in underparametrized Deep Neural Networks]{Subaging in underparametrized Deep Neural Networks}

\author{Carolina Herrera Segura}
\address{Instituto de Física, Universidad de Antioquia, Colombia~\footnote{carolina.herreras@udea.edu.co}}

\author{Edison Montoya}
\address{BCFort~\footnote{Regular Instructor at {\it Universidad de Antioquia}, and currently CEO of \hyperlink{www.bcfort.com}{BCFort}, a company of Blockchain and Artificial Intelligence}, \footnote{edison@bcfort.com}}

\author{Diego Tapias}
\address{Institute for Theoretical Physics, University of G\"ottingen, Germany \footnote{ diego.tapias@theorie.physik.uni-goettingen.de}}

\begin{abstract}
  %  {In this work, we consider a simple classification problem to show that the interplay between landscape and (learning) dynamics in the case of finite Deep Neural Networks in the underparametrized regime gives rise to effects similar to those associated with glassy systems. {In agreement with recent findings, we identify two regimes during training: one of fast learning followed by one with slow dynamics that exhibits subaging and is connected to a dynamical exploration of the bottom of the landscape of the loss function}. We show that our results are generic to different architectures and maintained in the more complex scenario of the MNIST database.}
  
{We consider a simple classification problem to show that the dynamics of finite--width Deep Neural Networks in the underparametrized regime gives rise to effects similar to those associated with glassy systems, namely a slow evolution of the loss function and aging. Remarkably, the aging is sublinear in the waiting time (subaging) and the power--law exponent characterizing it is robust to different architectures under the constraint of a constant total number of parameters. Our results are maintained in the more complex scenario of the MNIST database.  {We find that for this database there is a unique exponent ruling the subaging behavior in the whole phase.}}

\end{abstract}

%\pacs{{{XXXX}}}

%%%%%%%%%%%%%%%%%%%%%%%%%%%%%%%%%%%%%%%%%%%%%%%%%%%%%%%%%%%%%%%%%%
\section{Introduction}

Understanding the (supervised) learning of Deep Neural Networks (DNNs) continues to be one of the biggest challenges from a theoretical point of view~\cite{carleo2019machine}. Despite the existence of specific cases in which a significant understanding has been reached, such as the perceptron~\cite{seung1992statistical, franz2016simplest},  the single hidden layer in the mean field limit~\cite{mei2018mean} and the infinite--width limit~\cite{jacot2018neural, geiger2021landscape} (see also~\cite{carleo2019machine}),  in the more realistic case of finite--width and a finite number of nodes, a deeper understanding is still demanded. Steps toward this direction require the analysis of the interplay between the landscape of a generic Deep Neural Network and the (learning) dynamics. 

%{For a given DNN and a dataset, it is non-trivial to tell the generic properties of the corresponding landscape, i.e. the number of critical points for a given value of the loss (a.k.a. complexity, see~\cite{auffinger2013random, ros2019complexity} for a definition in the context of spin glasses, see also~\cite{choromanska2015loss} for results on this matter for a simple model of a DNN).The dataset plays the role of (quenched) disorder~\cite{Baity_Jesi_2019} and is an important piece in the understanding of the learning.  }
The landscape refers to the loss (hyper)--surface generated by the parameters of a network (weights and biases) given a dataset, and its generic properties are crucial in the understanding of learning. As a scalar valued function, it is characterized by the existence of multiple critical points, i.e. local minima, saddles and local maxima. It is a non--trivial problem, however, to find them as a function of the loss for a generic DNN (see~\cite{choromanska2015loss} for results on this matter for a simple model). For the specific architectures/datasets with a small number of parameters considered in this work, the local minima and saddles of index one connecting them have been reported  and visualized with disconnectivity graphs in~\cite{Verpoort2020}.
%see ~\cite{auffinger2013random, ros2019complexity} for a definition in the context of spin glasses, see also~\cite{choromanska2015loss} for results on this matter for a simple model of a DNN). For specific architectures/datasets with a small number of parameters considered in this work, the local minima and saddles of index-1 connecting them have been reported  and visualized with disconnectivity graphs in reference~\cite{Verpoort2020}.
%Ideally, a landscape characterization analogous to what has been done for the spherical $p$--spin is prosecuted~\cite{auffinger2013random, ros2019complexity}. Unfortunately, the reality is still far from this scenario.

Apart from the landscape, the other relevant piece in the puzzle of (supervised) learning is the role of the (microscopic) dynamics. In practical applications, \emph{Stochastic Gradient Descent} (SGD) continues to be one of the most popular optimization algorithms~\cite{mehta2019high} and will be the focus of our work. The effects of SGD are tracked through the observation of global quantities, the loss function being by far the most significant quantity of interest. In addition, it has been shown that a two--time observable as the mean square displacement (MSD) in the parameter space is helpful to reveal additional features hidden in the loss evolution~\cite{Baity_Jesi_2019, kunin2021limiting, chen2022anomalous}. 

In this work we examine the interplay between landscape and dynamics by implementing SGD for some of the architectures/datasets--size considered in~\cite{Verpoort2020} {(in the context of structural glasses, the interplay between landscape and dynamics has been studied in detail, for instance, in references~\cite{de2008energy, niblett2016dynamics, niblett2017pathways})}. We observe that the details of the corresponding landscapes are immaterial for the evolution of global functions such as the loss or the MSD, which gives hints of how ``real--life'' networks learn. As a matter of fact, we find that the similarities  across architectures/landscapes is not only qualitative but also quantitative as the (sub)aging exponents support. To be more specific, we show that despite the differences in the corresponding landscapes, the dynamical behavior of DNNs in the underparametrized regime is glassy, in the sense that the long--time dynamics is slow, time--translation invariance is broken and the MSD depends on the ``age'' of the system~\cite{fielding2000aging,Baity_Jesi_2019,arceri2020glasses}. Additionally, we show that the type of aging is sublinear in the age and that the power--law exponent characterizing it is independent of the architecture if the ratio $\alpha = P/N$, with $P$ the number of parameters and $N$ the number of data, is constant. {We also show analogous results using the MNIST database~\cite{lecun1998mnist} to support the generality of our conclusions. {In particular, for this database, we obtain numerically the unique exponent characterizing the subaging across the whole regime.}}
%A practical way of differentiating ``real-life'' networks relies on the ratio of the number of parameters with respect to the number of data. 

We focus here on networks with a small number of parameters with respect to the number of data, i.e.~underparametrized, as they are sufficiently good for the classification problem that we have on hand (LJAT19 database~\cite{ljat19data}), and also because for this regime a correspondence with glassy features has already been pointed out~\cite{Baity_Jesi_2019}.

The manuscript is structured as follows: In section~\ref{landsc} we introduce the LJAT19 database together with the relevant information regarding the landscape structure; moreover, we discuss the dynamical functions to be considered. In section~\ref{results} we provide our main results and discuss them. Finally, in section~\ref{conc} we draw the conclusions of our analysis.

%%%%%%%%%%%%%%%%%%%%%%%%%%%%%%%%%%%%%%%%%%%%%%%%%%%%%%%%%%%%%%%%%%
\section{Landscape structure and dynamics}
\label{landsc}

We consider a simple classification problem as originally introduced in~\cite{ballard2016energy}. This refers to the database LJAT19~\cite{ljat19data}, which consists of data with three coordinates (input) that describe a triatomic system and a label (output) that corresponds to one out of four possible stable configurations that are reached using standard molecular dynamics. In~\cite{Verpoort2020}, the difficulty of the classification problem was increased by reducing to two the input coordinates, with the  implication that the maximum possible accuracy during training, defined as the ratio of correctly classified samples over the total number of samples, is around $0.8$.

In~\cite{Verpoort2020}, the landscape of small fully--connected DNNs is studied as the depth (number of hidden layers, $H = 1, 2, 3$) and dataset size ($N$) is varied, while keeping the number of parameters low and approximately constant (see details in appendix~\ref{methods}). The idea behind this procedure is to keep the dimension of the parameter space approximately equal and, in this way, make a meaningful comparison of architectures.  

The loss function to be considered is the cross entropy with softmax outputs and a $L2$ regularization, normalized by the number of data. This is:

{\begin{equation}
    \mathcal{L}_{D}(\bm{\theta}) = - \frac{1}{N} \sum_{n=1}^{N} \ln \left( 
    \frac{{\rm{e}}^{a_n^{(k)}({\bm{\theta}})}}{\sum_{k=1}^K {\rm{e}}^{a_n^{(k)} ({\bm{\theta}}) } } \right) + \lambda ||\bm{\theta}||^2  \, ,
\end{equation}}
{where $D$ refers to the dataset, ${\bm{\theta}} = (\theta_1, \ldots, \theta_P)$ is the vector of all the weights (including biases), $a_n^{(k)}$ refers to the softmax output for the true $k$-th class associated with the $n$-th data and $\lambda$ is the regularization constant.}
 
Specified the dataset $D$ and the architecture of the network, the set of local minima $\{{\bm{\theta}}^*\}$ of  the function $\mathcal{L}_{D}(\bm{\theta})$ can be extracted with a basin--hopping algorithm~\cite{wales2006potential}, and the transition states {(saddles of index-1)} connecting those minima are searched with Discrete Path Sampling~\cite{wales2006potential}. For the LJAT19 database, this characterization was done in reference~\cite{Verpoort2020} and the results unveil how the structure of the landscape is database size dependent. In particular they show that for a small number of data $N$, the multilayer architectures present many competing low-lying minima with similar loss values, whereas the single hidden layer landscape is funneled with a small number of local minima; on the other hand, in the regime of high $N$, all the landscapes turn out to be funneled with a small number of local (shallow) minima (see~\cite{wales1998archetypal} for a visualization of funneled landscapes).

In this manuscript, unless otherwise explicitly stated, we show results for the small dataset regime $N = 1000$. To help the reader to have an idea of the complexity of the landscape, we report here the order of the number of local minima found ($n_{\rm{min}}$) for the considered dataset (cf. Table 1 in~\cite{Verpoort2020}): single--hidden layer, $n_{\rm{min}} \sim 10$, two--hidden layer, $n_{\rm{min}} \sim 10^3$, three--hidden layer, $n_{\rm{min}} \sim  10^4$. In all the cases, the loss value of the deepest found minimum is around $0.5$.

Once the landscape is defined, the next step is to perform the optimization dynamics that allows its exploration. We implement the standard Stochastic Gradient Descent (SGD) with a fixed learning rate and fixed batch size (details in appendix~\ref{methods}. We remark that the validity of our results is independent from the chosen values). Then we keep track of the evolution of the loss function during training together with the mean square displacement (MSD). This is defined between times $t_{\rm{w}}$ and $t_{\rm{w}} + t$ as~\cite{Baity_Jesi_2019}
\begin{equation}
    \Delta (t_{\rm{w}}, t_{\rm{w}} +t) = \frac{1}{P} \sum_{i=1}^P (\theta_i(t_{\rm{w}}) - \theta_i(t_{\rm{w}} + t))^2
    \label{eq:MSD}
\end{equation}
with $t_{\rm{w}}$ the so-called ``waiting time'' and $P$ the total number of parameters. The function~(\ref{eq:MSD}) characterizes the overall motion between two times in the parameter space.

{In systems with glassy dynamics, a way of characterizing the slowness of the evolution and in particular the \emph{aging} is via the scaling of a characteristic time $t^*$ with $t_{\rm{w}}$~\cite{berthier2000sub, arceri2020glasses}. As a matter of fact, $t^*$ may be extracted by fixing a value of $\Delta = \Delta^*$ and then solving for $t$ as a function of the waiting time. In particular, the scaling $t^* \sim t_{\rm{w}}^\mu$ with $\mu$ a positive number is quite generic in glassy models~\cite{bouchaud1999aging, rinn2000multiple, berthier2000sub}. Case $\mu =1$ is denoted as simple aging, whereas subaging and superaging denote the cases $\mu <1$ and $\mu >1$, respectively. We obtain numerically $t^*(t_{\rm{w}})$ for the architectures/datasets discussed in this work and show that a subaging scenario is compatible with the data.}

Finally, we characterize the local geometry of the landscape along the evolution by looking at the distribution of the Hessian eigenvalues of the loss function (as proposed in~\cite{sagun2017empirical}), i.e. $H^{D}_{ij}({\bm{\theta}}) = \frac{\partial^2 \mathcal{L}_D}{\partial \theta_i \partial \theta_j}$. As the stochastic evolution is driven towards local minima, the number of negative eigenvalues is highly reduced. However, because of the stochastic evolution, the dynamics do not get trapped in the local minima and negative curvatures prevail to some extent~\cite{wei2019noise}. In the next section, we illustrate that the change in the distribution helps to gain intuition about the training.

%As a summary, we consider the database LJAT19 together with the architectures and the associated loss landscapes already analyzed in~\cite{Verpoort2020}. Then we train the networks with the standard SGD, and across the evolution we track the loss function and the mean square displacement. Finally, we examine the local geometry by looking at the distribution of the Hessian eigenvalues.

%\begin{figure}[t]
%\centering
%\includegraphics[width=0.75\textwidth]{figures/archetypal_discgraph.png}
%\caption{Disconnetivity grapths for the architectures with one, two and three hidden layers and a training databse of 1000 points from Verpoort et al. \cite{Verpoort2020}. %Allí se reportan 7 mínimos y 33 estados de transición para la arquitectura de una capa oculta, {13336} mínimos y {41837} estados de transición para la de dos capas ocultas y {85150} mínimos y {263615} estado de transición para la de tres capas ocultas.
%}
%\label{fig:archdisc}
%\end{figure}

%%%%%%%%%%%%%%%%%%%%%%%%%%%%%%%%%%%%%%%%%%%%%%%%%%%%%%%%%%%%%%%%%%
\section{Results}
\label{results}
%\subsection{Dynamics}
\begin{figure}[t]
     \centering
     \begin{subfigure}[b]{0.45\textwidth}
         \centering
         \includegraphics[width=\textwidth]{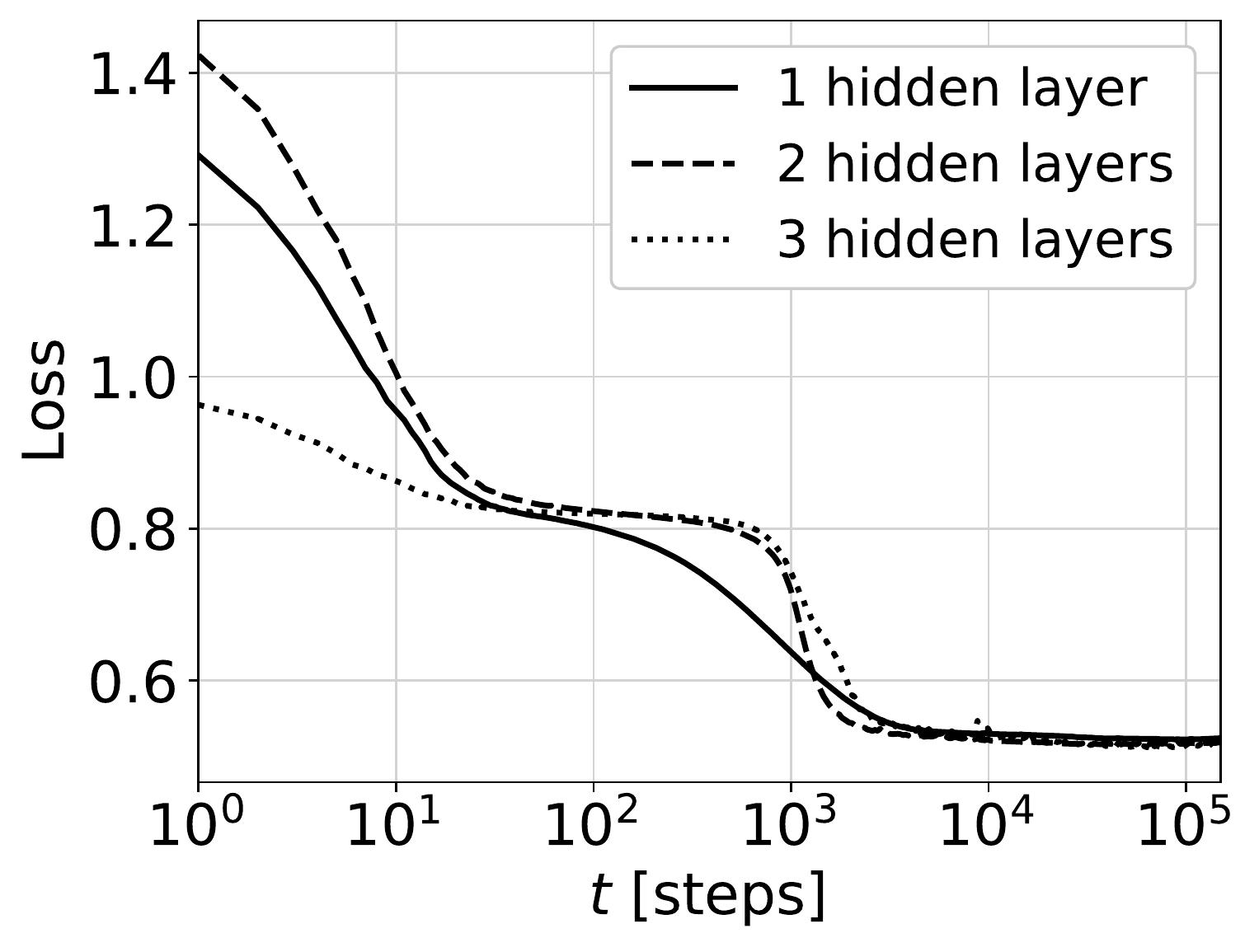}
         \caption{}
         \label{fig:70loss}
     \end{subfigure}
     \hfill
     \begin{subfigure}[b]{0.45\textwidth}
         \centering
         \includegraphics[width=\textwidth]{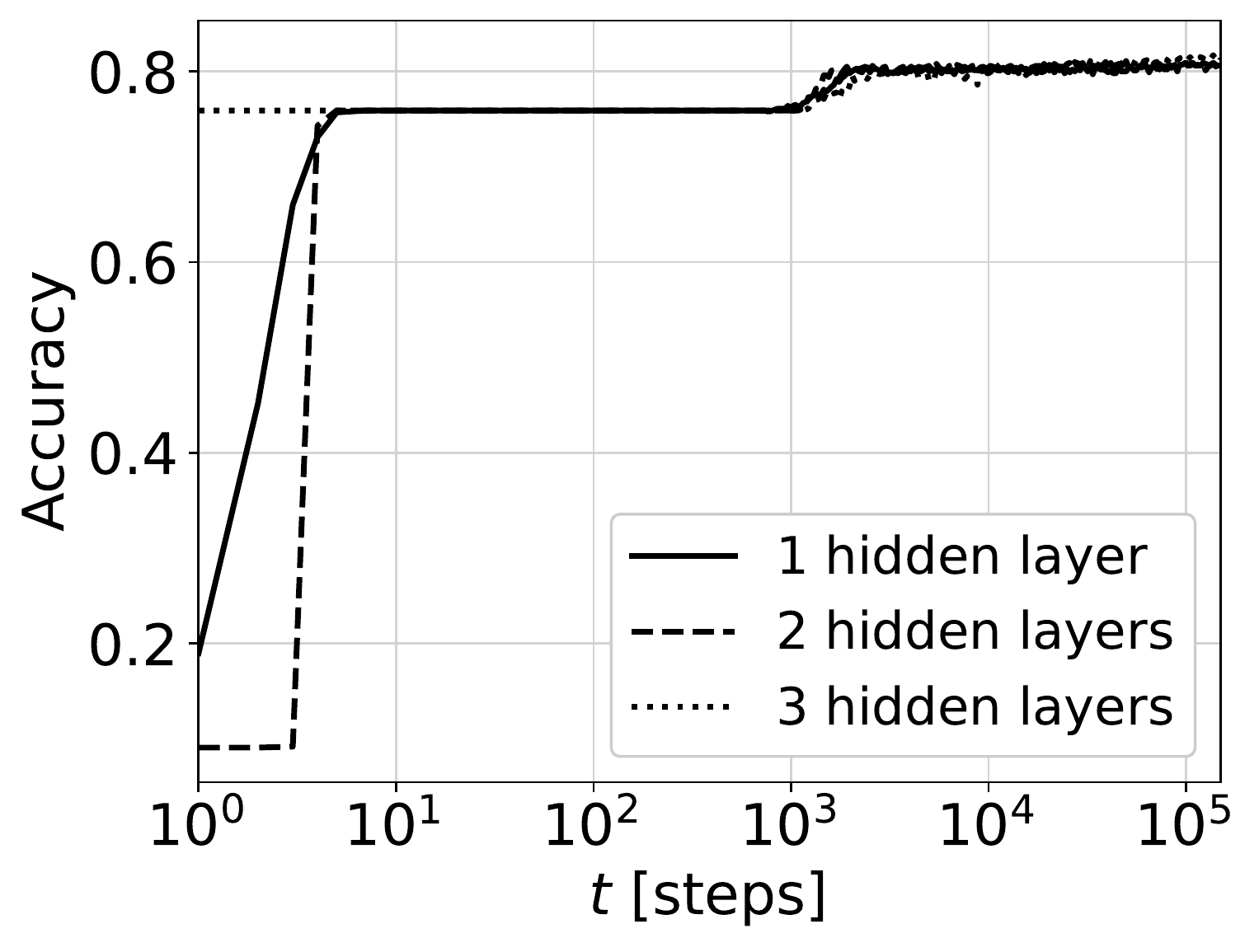}
         \caption{}
         \label{fig:70acc}
     \end{subfigure}
        \caption{Loss function (left) and accuracy (right) in terms of the number of steps of the SGD algorithm for each architecture and dataset size $N = 1000$. The solid line corresponds to one hidden layer, the discontinuous line to two hidden layers, and the dotted line to three hidden layers. We consider {15000} epochs with a batch size of 100 that leads to {150000} steps.}
        \label{fig:loss70}
\end{figure}

\begin{figure}[t]
     \centering
     \begin{subfigure}{0.45\textwidth}
         \centering
         \includegraphics[width=\textwidth]{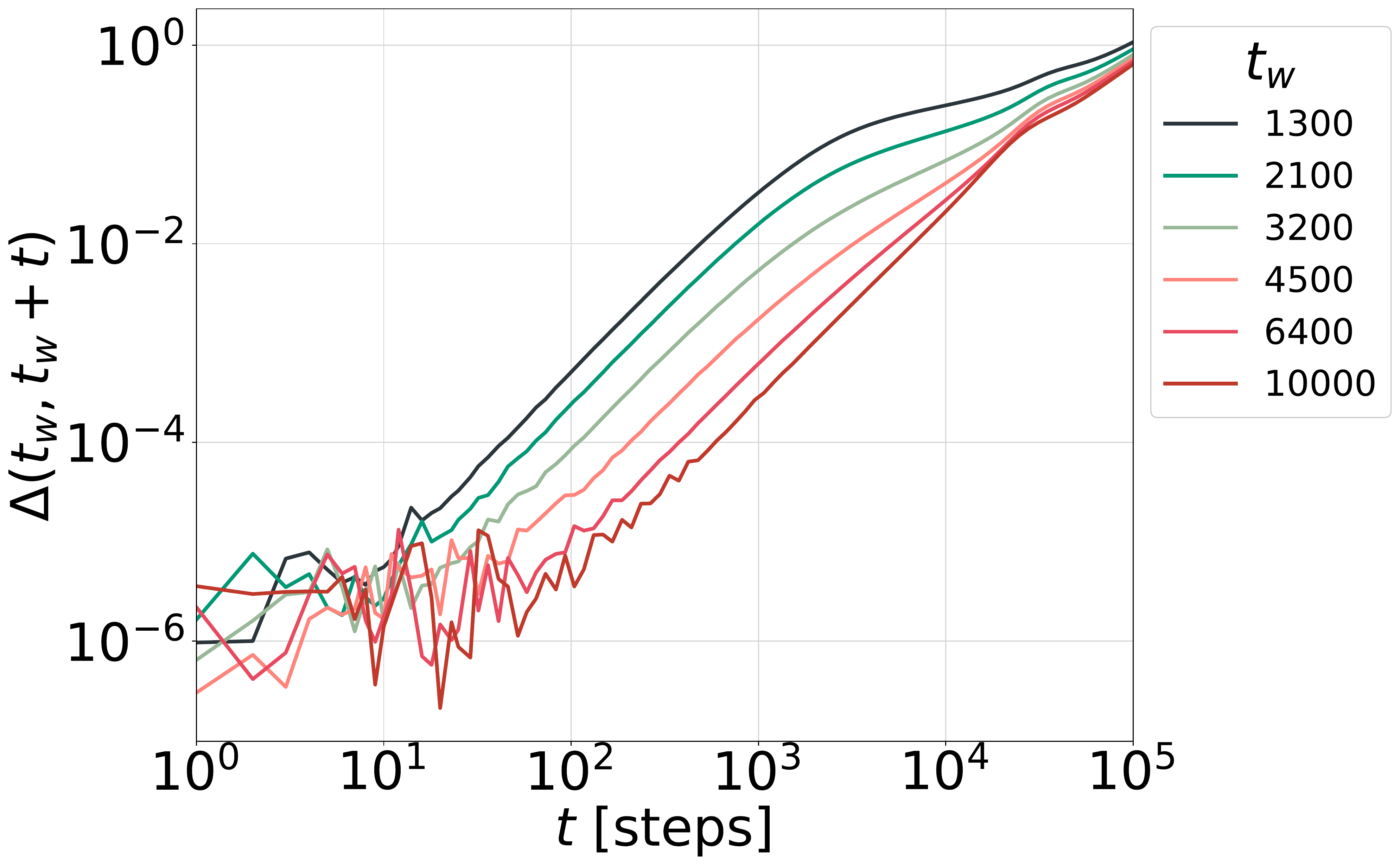}
         \caption{}
         \label{fig:corr701H}
     \end{subfigure}
     \hfill
     \begin{subfigure}{0.45\textwidth}
         \centering
         \includegraphics[width=\textwidth]{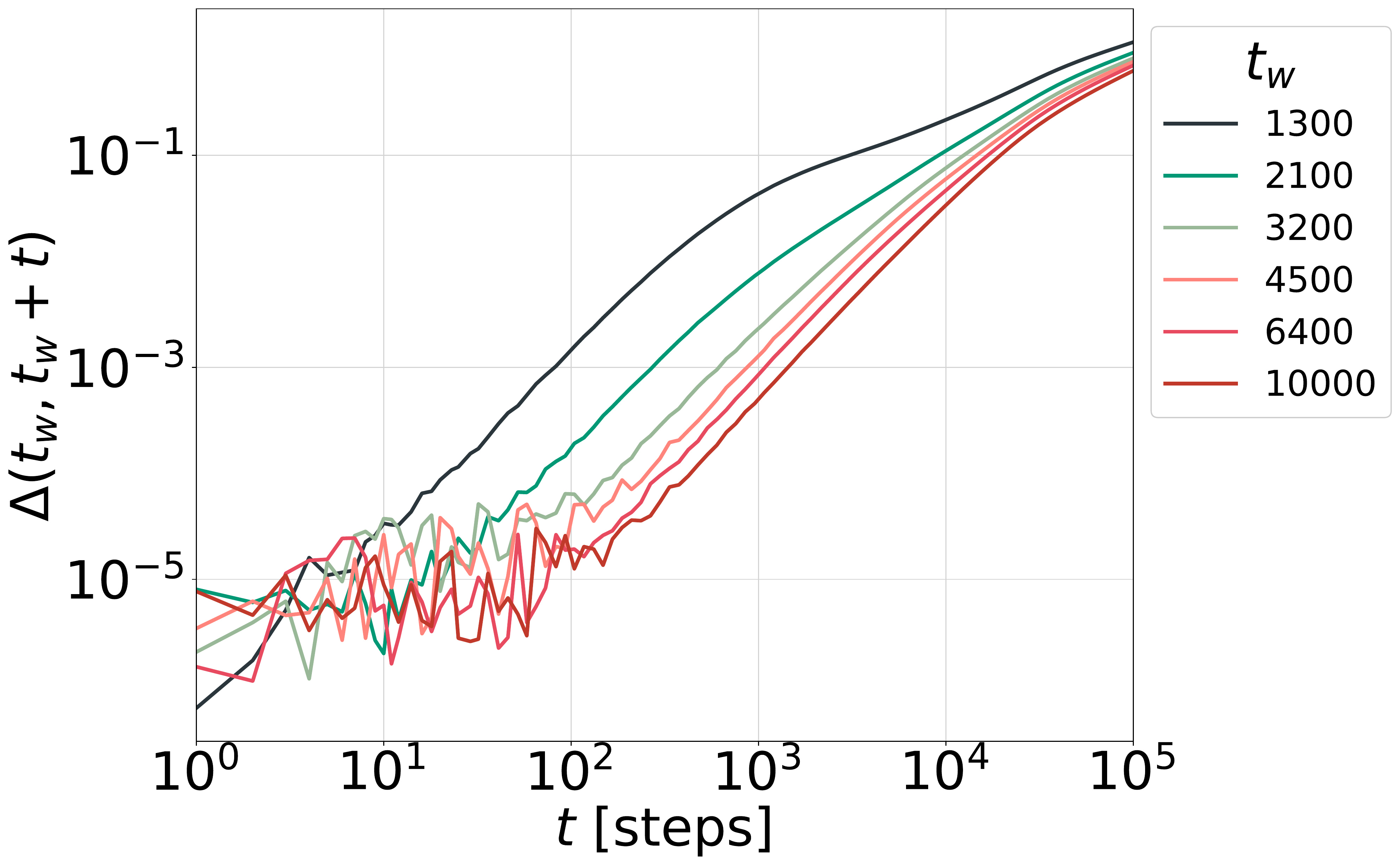}
         \caption{}
         \label{fig:corr702H}
     \end{subfigure}
     \hfill
       \begin{subfigure}{0.45\textwidth}
         \centering
         \includegraphics[width=\textwidth]{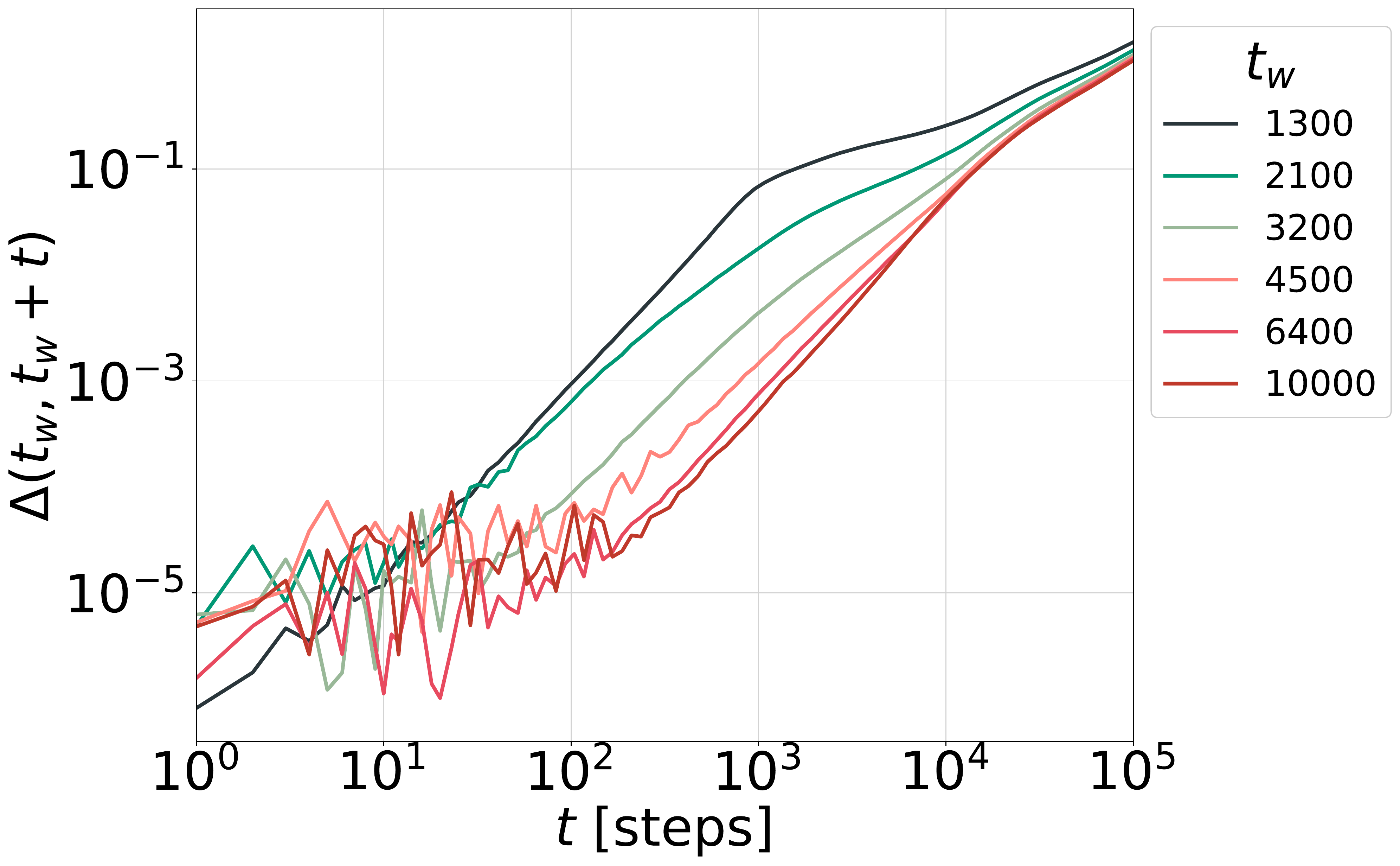}
         \caption{}
         \label{fig:corr703H}
     \end{subfigure}
     \hfill
       \begin{subfigure}{0.44\textwidth}
         \centering
         \includegraphics[width=\textwidth]{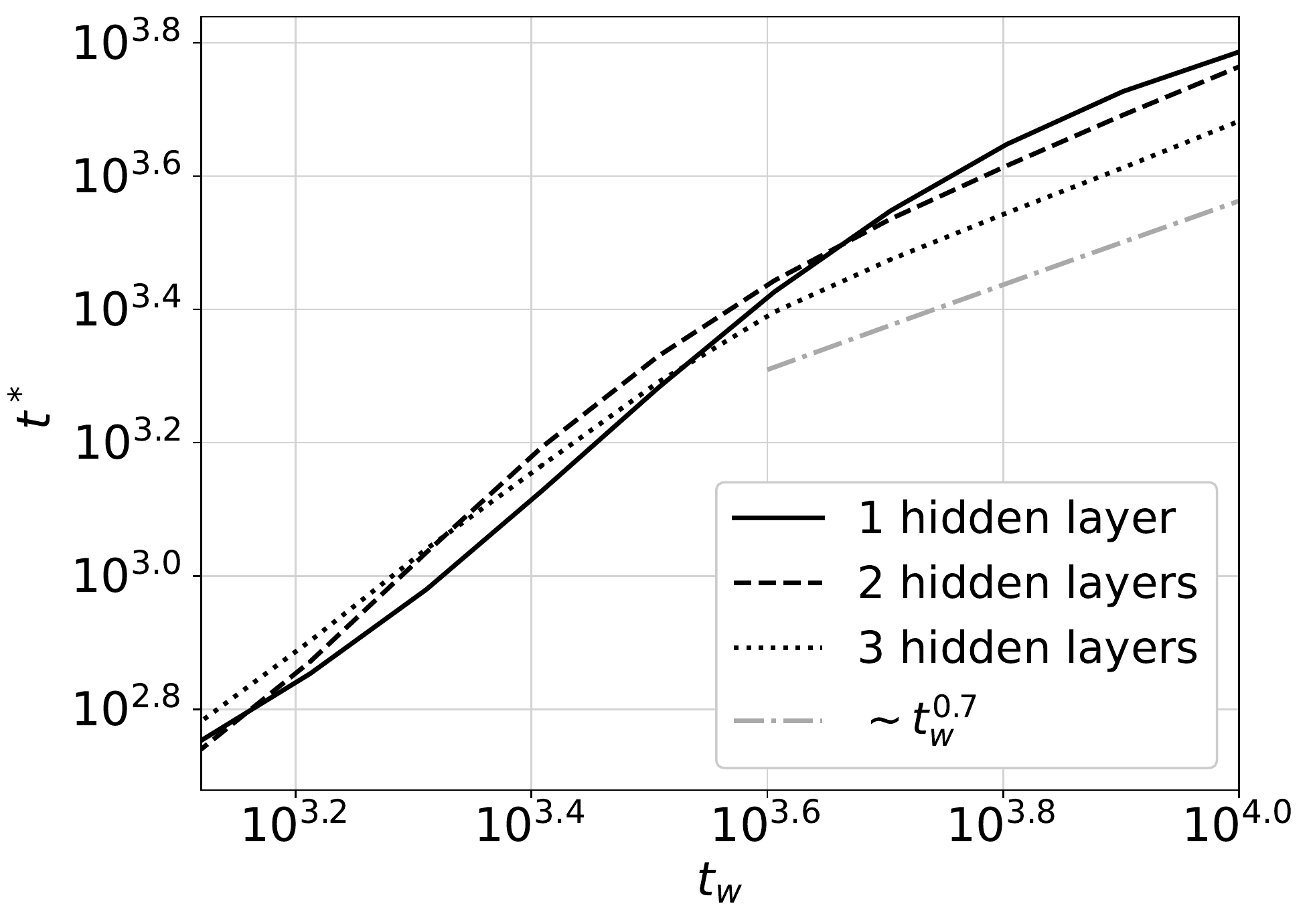}
         \caption{}
         \label{fig:twtljat19}
     \end{subfigure}
    
        \caption{Mean square displacement (eq.~(\ref{eq:MSD})) for the same data shown in Fig.~\ref{fig:loss70} evaluated at waiting times $t_{\rm{w}} > 10^3$, where the system has entered (or is close to enter) in the slow learning phase. Figure {\bf (a)} corresponds to the architecture with one hidden layer, {\bf (b)} to two hidden layers and {\bf (c)} to three hidden layers. Figure {\bf (d)} shows the characteristic time $t^*$ for each $t_{\rm{w}}$; for this plot, each curve is an average over 100 independent runs. The scaling with $t_{\rm{w}}$ shown as a dot-dashed line is set by hand with the purpose of illustration.}
        \label{fig:corr70}
\end{figure}

%%@@Shall we mention why steps and no epochs?
We trained networks with one, two and three hidden layers with approximately {$P \approx 70$} each using the database LJAT19 (details in appendix~\ref{methods}). We show the train loss function and accuracy against the number of steps in figure~\ref{fig:loss70}; a step refers to one iteration across a single minibatch.  As a general observation, we see that for the three architectures both the loss and the accuracy reach similar values and they enter into the regime where the lowest local minima lie ($\mathcal{L} \approx 0.5$) at $t \approx 10^3$. Regarding the specific form of the loss evolution, which exhibits an intermediate plateau especially for the multilayer cases, we point out that the corresponding landscapes exhibit many local minima at values of $\mathcal{L} \approx 0.8$~\footnote{Obtained using GMIN~\cite{wales1999gmin} adapted to DNNs, see for instance~\cite{maxpylfl}.}, which explains this transient behavior. 

\begin{figure}[t]
     \centering
     \begin{subfigure}[b]{0.45\textwidth}
         \centering
         \includegraphics[width=\textwidth]{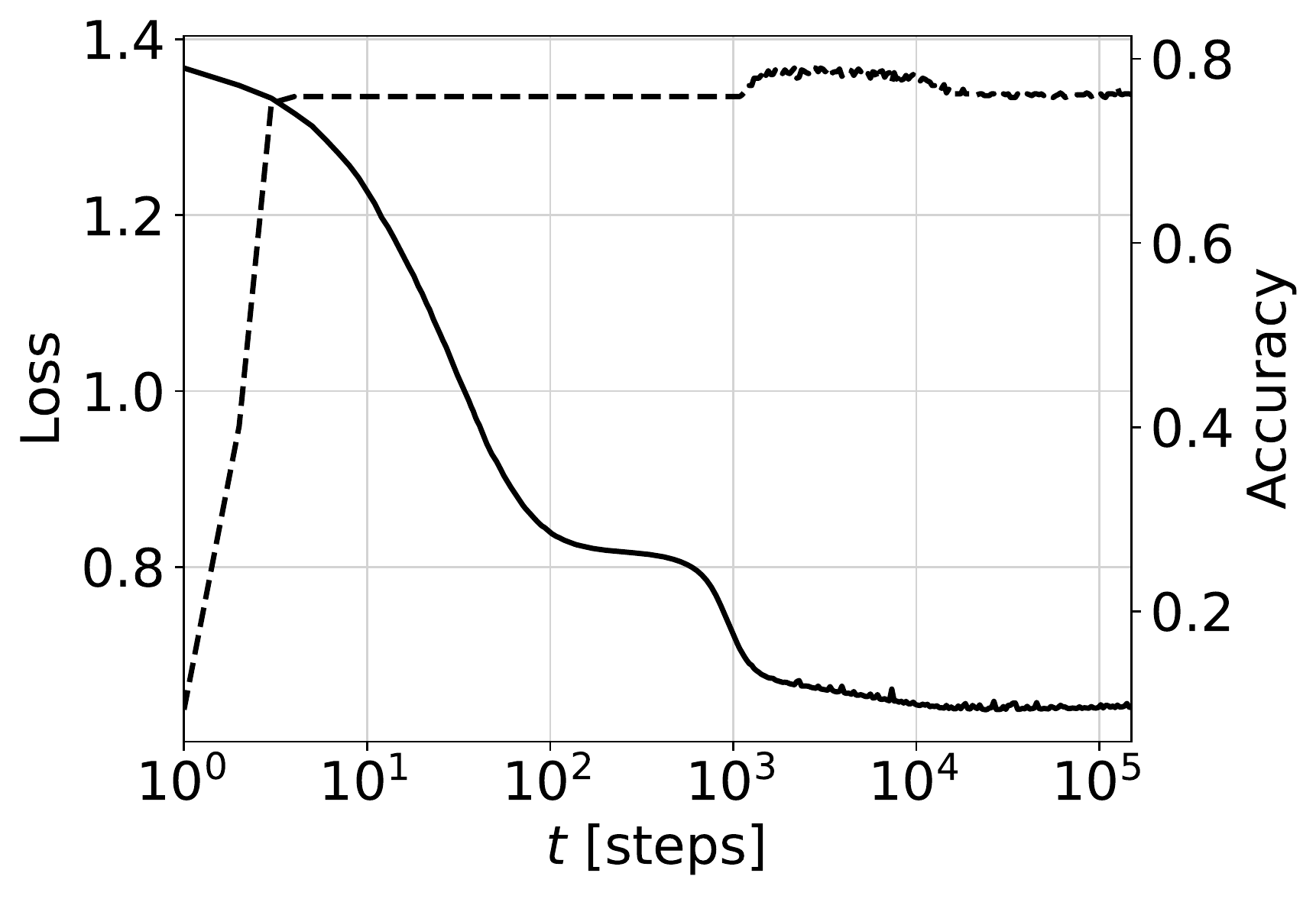}
         \caption{}
         \label{fig:minlossacc}
     \end{subfigure}
     \hfill
     \begin{subfigure}[b]{0.45\textwidth}
         \centering
         \includegraphics[width=\textwidth]{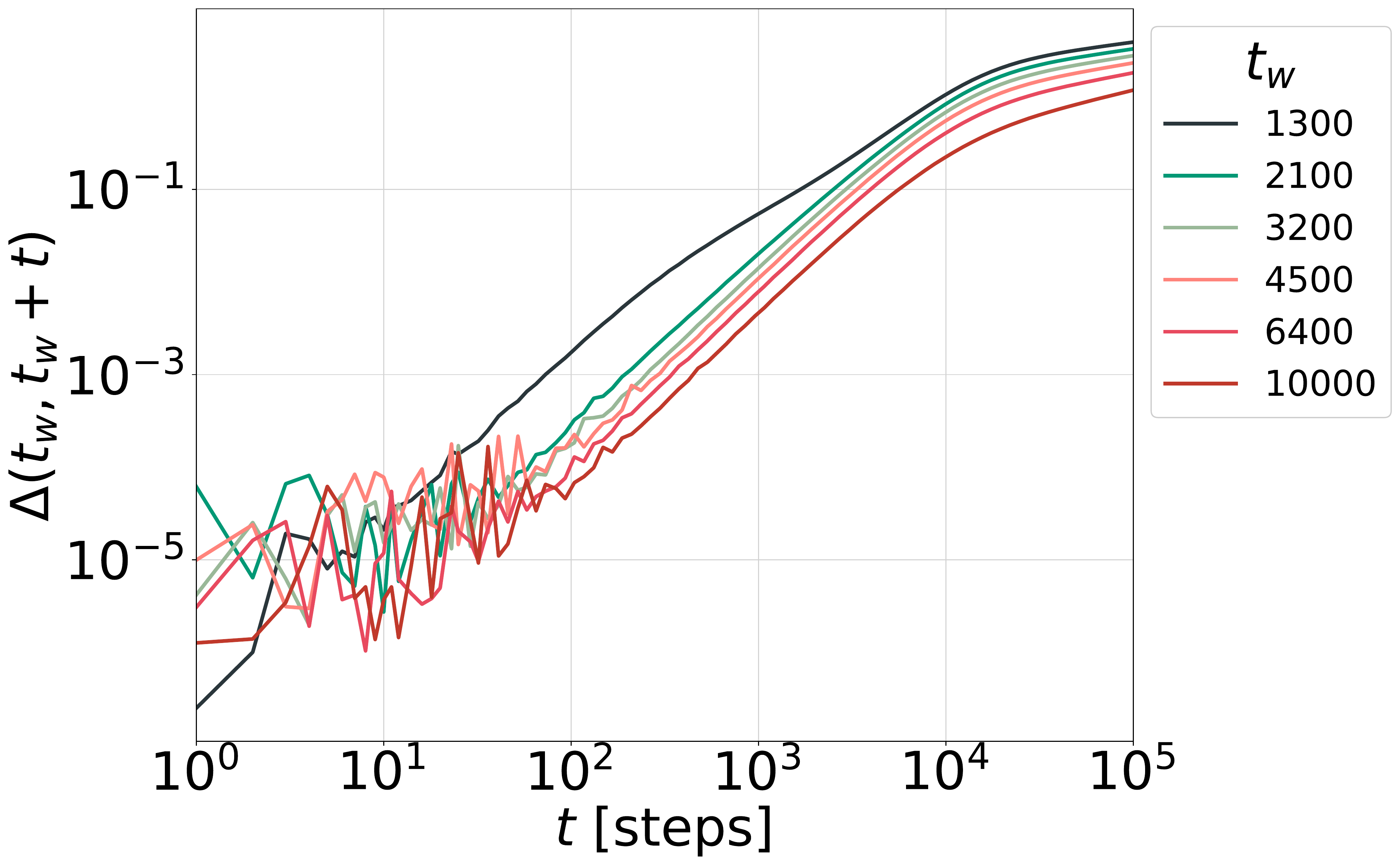}
         \caption{}
         \label{fig:mincorr}
     \end{subfigure}
     
        \caption{{\bf (a)} Loss function (left axis, solid line) and accuracy (right axis, discontinuous line) in terms of the number of steps of the SGD algorithm for three hidden layers with one neuron per layer. {\bf (b)} Mean square displacement for the same case evaluated at waiting times $t_{\rm{w}} > 10^3$.}
        \label{fig:lossmin}
\end{figure}

%@@Esta discusi\'on est\'a simplificada pero posiblemente mejor presentada ahora, ver unos p\'arrafos abajo.@@ In figure \ref{fig:corr70} we show the mean square displacement for each of the networks. In these we observe aging because of the dependence between the MSD and $t_w$. Figure \ref{fig:corr703H} shows interrupted aging, where the time scale doesn’t change for $t_w \geq 4500$. We can also divide this plot into two regions: one in which the MSD shows normal diffusion as it is proportional to $t$ ($t_w = 50, 320, 750$) and another one in which there is noise at the start of the curve that later becomes normal diffusion ($t_w \geq 1300$). Comparing to the loss function in figure \ref{fig:loss70}, we see the noise in the second region appears when the loss function arrives to a plateau. This could be due to the system arriving at the bottom of the landscape, where it starts a diffusion process given by the linear behaviour.

%@@De entrada vaamos a decir que es underparametreized@@ By modifying the original number of input coordinates, solving the LJAT19 problem completely is no longer possible and we can see this in the fact that the loss stalls at a value much higher than zero and the accuracy stays at around 0.8. This makes it hard to identify the regime to which this model belongs.

According to Figure~\ref{fig:loss70} and the information on the landscapes, we separate the evolution of the loss into two phases. In the first one, learning is relatively fast, despite the eventual existence of intermediate plateaus. In the second phase, learning is slow, and the system explores the bottom of the landscape composed of multiple local minima at around the same depth. This division is consistent with other descriptions, as for instance references~\cite{frankle2020early, fengPhases}. Following~\cite{fengPhases}, in the first phase, learning is fast, and the difference between consecutive gradient updates is aligned; whereas the second phase is an exploration phase, in which the difference between updates is distributed in all directions (see also~\cite{kunin2021limiting}). 

{We characterize further the slow learning phase with the mean square displacement. In this phase, there is a non-translational invariant time dependence of the MSD, as Fig.~\ref{fig:corr70} reveals. According to this figure we can say that the overall motion becomes slower with the ``age''   $t_{\rm{w}}$ of the system as it happens for structural glasses~\cite{scalliet2019rejuvenation, arceri2020glasses}. This is the situation regardless of the architecture. To substantiate this, we compute numerically the function $t^*(t_{\rm{w}})$ obtained by fixing $\Delta^* = 10^{-2}$ and the result is shown in Fig.~\ref{fig:twtljat19}. Remarkably, the asymptotic scaling is independent of the architecture and reveals a subaging behavior.}

Few comments on this result are in order. Firstly, even though the value for $\Delta^*$ was chosen arbitrarily, Figure~\ref{fig:corr70} (a)--(c) shows that there exists an interval $\Delta^* \in [\Delta_{\min}, \Delta_{\max}]$ such that the scaling exponent $t^* \sim t_{\rm{w}}^\mu$ is independent or weakly dependent from $\Delta^*$. This can be read as  the vertical range in which the curves for different waiting times remain parallel.
%An alternative way to characterize this regime is by fitting the curves to the scaling $\Delta \sim D(t_w) t^\alpha$  and compute how the (generalized) diffusion coefficient $D$ changes with $t_{\rm{w}}$. 
Secondly, we see that for long relative times $t$, the different MSD curves tend to approach together; this indicates a change in the nature of aging for long times.
%and may also be characterized by fixing $\Delta^*$ to a higher value or finding the generalized diffusion coefficient in this second regime. 
This regime has not analogue in standard models of glassy systems. Its full characterization is left for future work.

%and suggests that in the long--time limit there is a unique relevant timescale. 
%. If we collected data for larger $t_{\rm{w}}$ we would observe that at certain point, the subaging exponent would start to decrease ($t^*$ would not change much with $t_{\rm{w}}$) and eventually it would reach zero. The situation is analogous to the observed for a binary Lennard--Jones glass after a quench at temperatures where the system can equilibrate but is close to the glass transition temperature~\cite{warren2013quench}. 

We also comment on the crossover around $t_{\rm{w}} \approx 10^{3.6}$ in Figure~\ref{fig:twtljat19}. This time is related to the overcoming of the regions with high loss values that lead to intermediate plateaus in the evolution of the loss function (Fig.~\ref{fig:70loss}) and is characteristic of the LJAT19 database. 

\begin{figure}[t]
     \centering
     \begin{subfigure}[b]{0.45\textwidth}
         \centering
         \includegraphics[width=\textwidth]{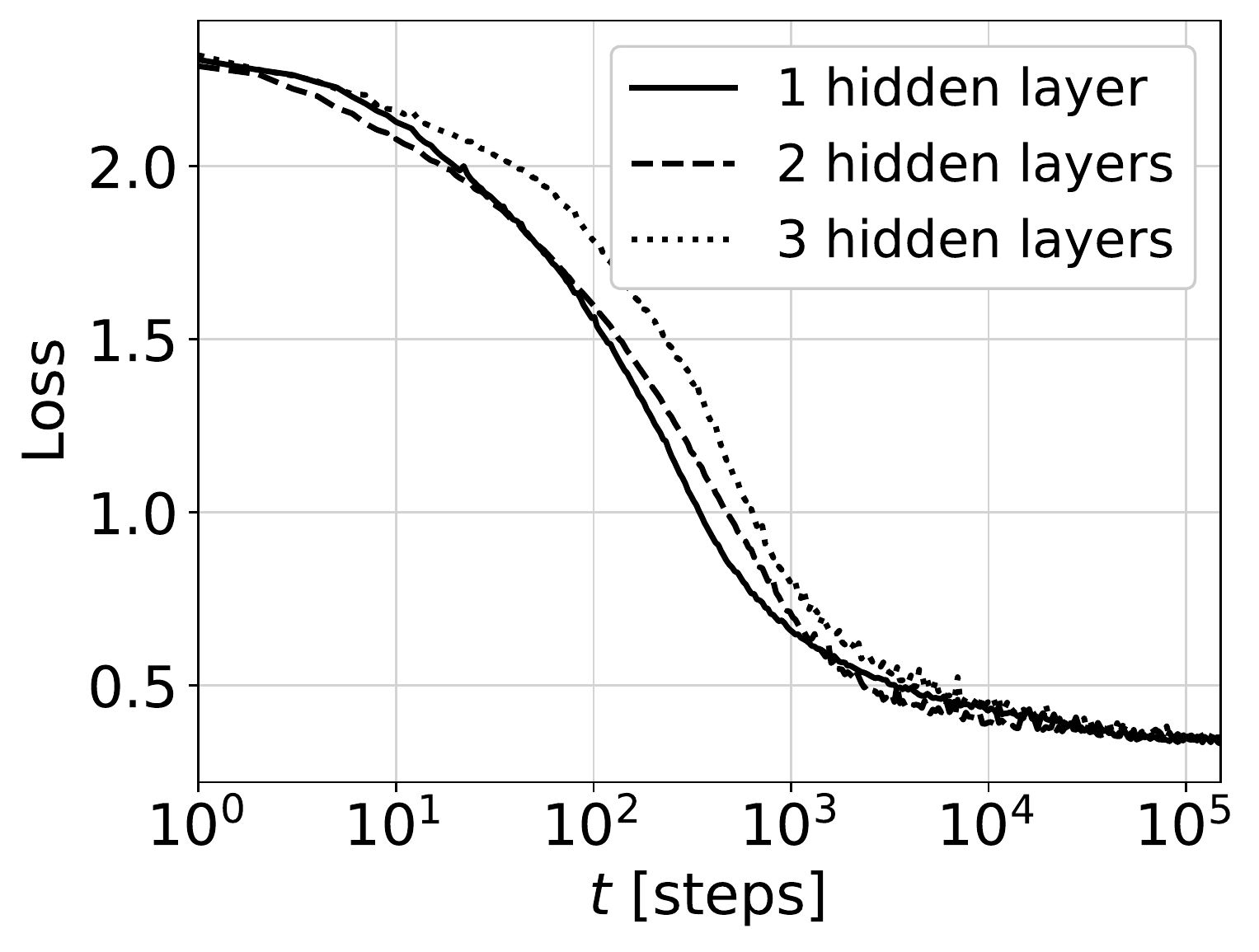}
         \caption{}
         \label{fig:lossHmnist}
     \end{subfigure}
     \hfill
     \begin{subfigure}[b]{0.45\textwidth}
         \centering
         \includegraphics[width=\textwidth]{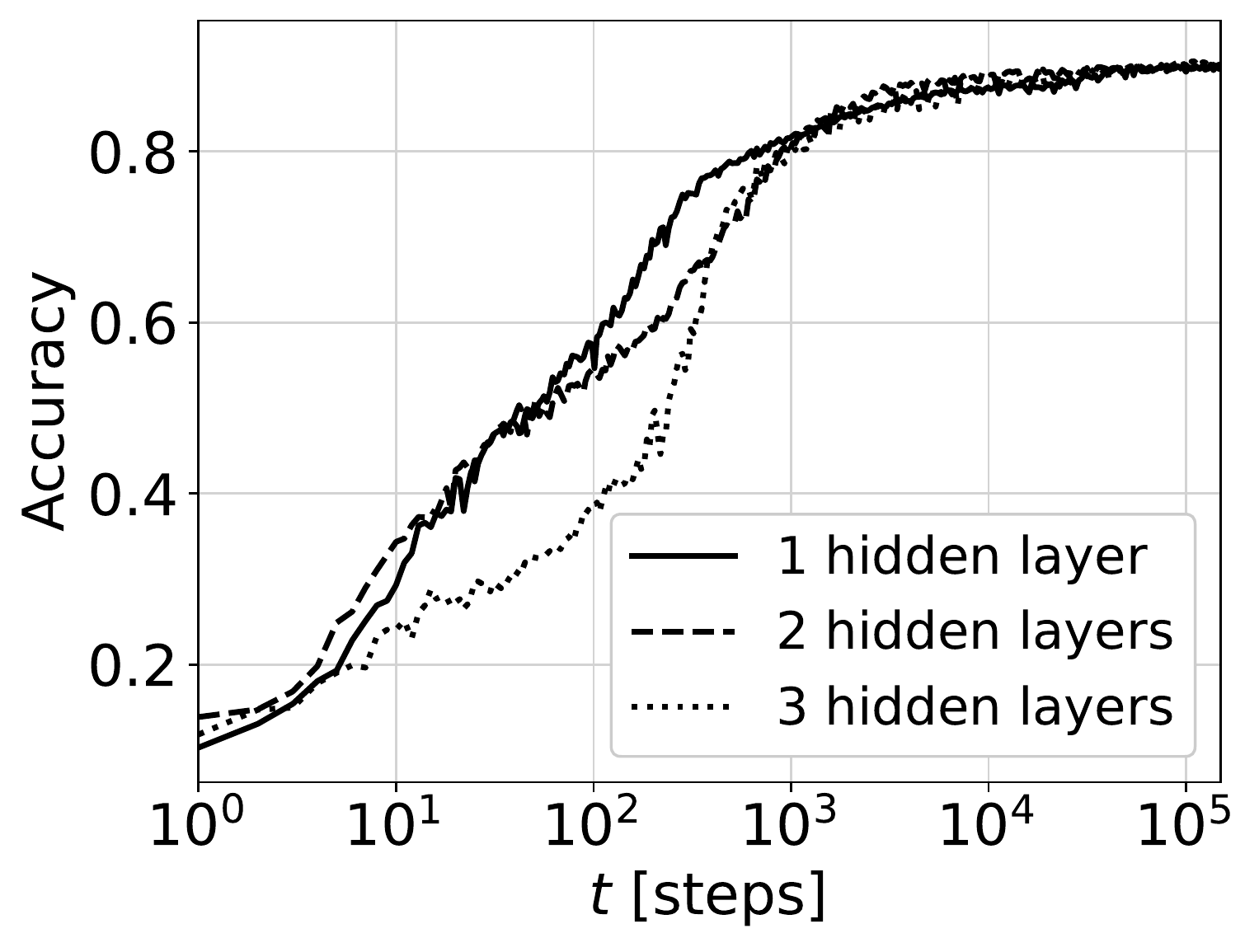}
         \caption{}
         \label{fig:accsHmnist}
     \end{subfigure}
        \caption{Loss function (left) and accuracy (right) in terms of the number of steps of the SGD algorithm for each architecture and the MNIST database.}
        \label{fig:lossmnist}
\end{figure}

\begin{figure}[b]
     \centering
     \begin{subfigure}{0.47\textwidth}
         \centering
         \includegraphics[width=\textwidth]{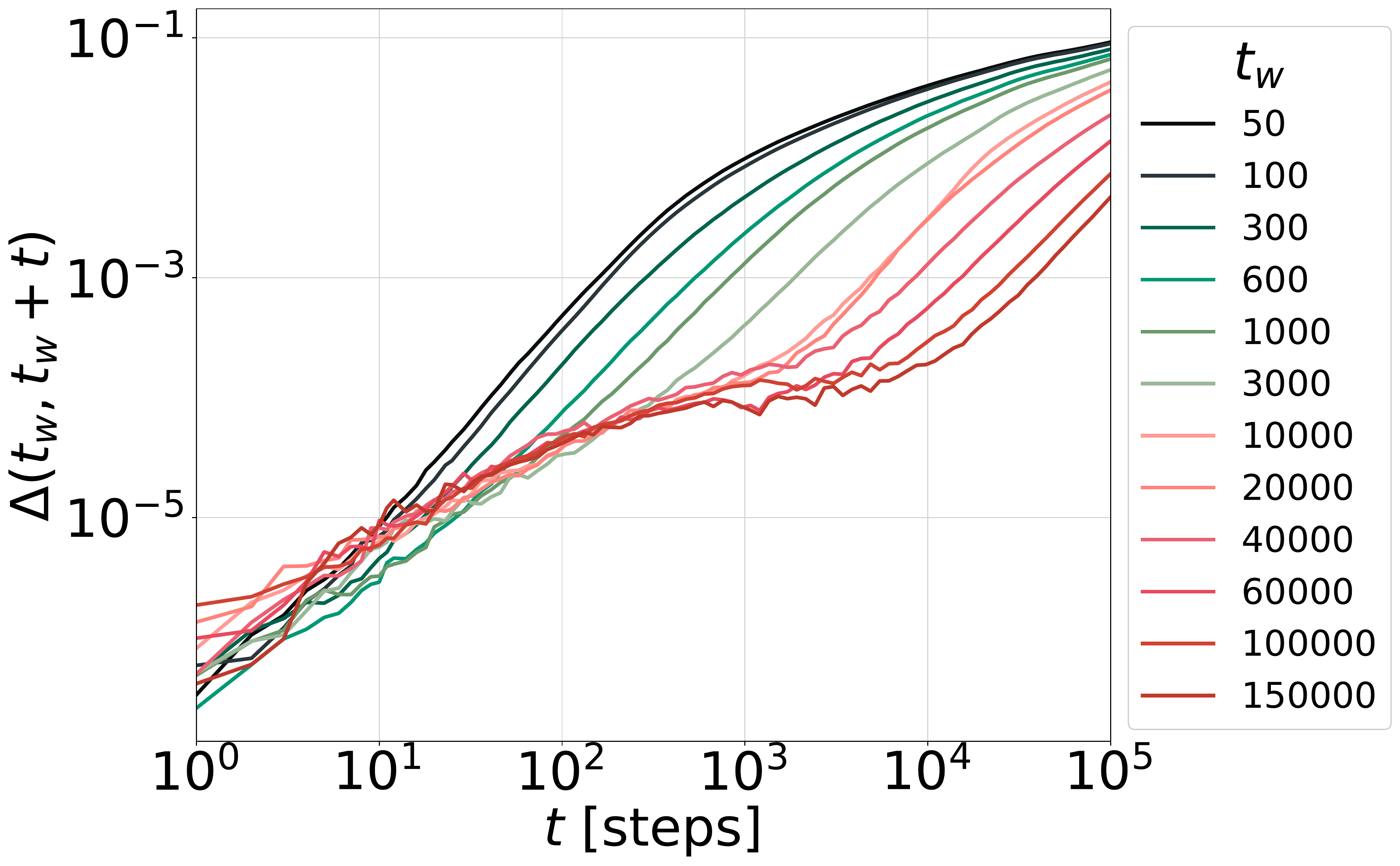}
         \caption{}
         \label{fig:corr1H}
     \end{subfigure}
     \hfill
     \begin{subfigure}{0.47\textwidth}
         \centering
         \includegraphics[width=\textwidth]{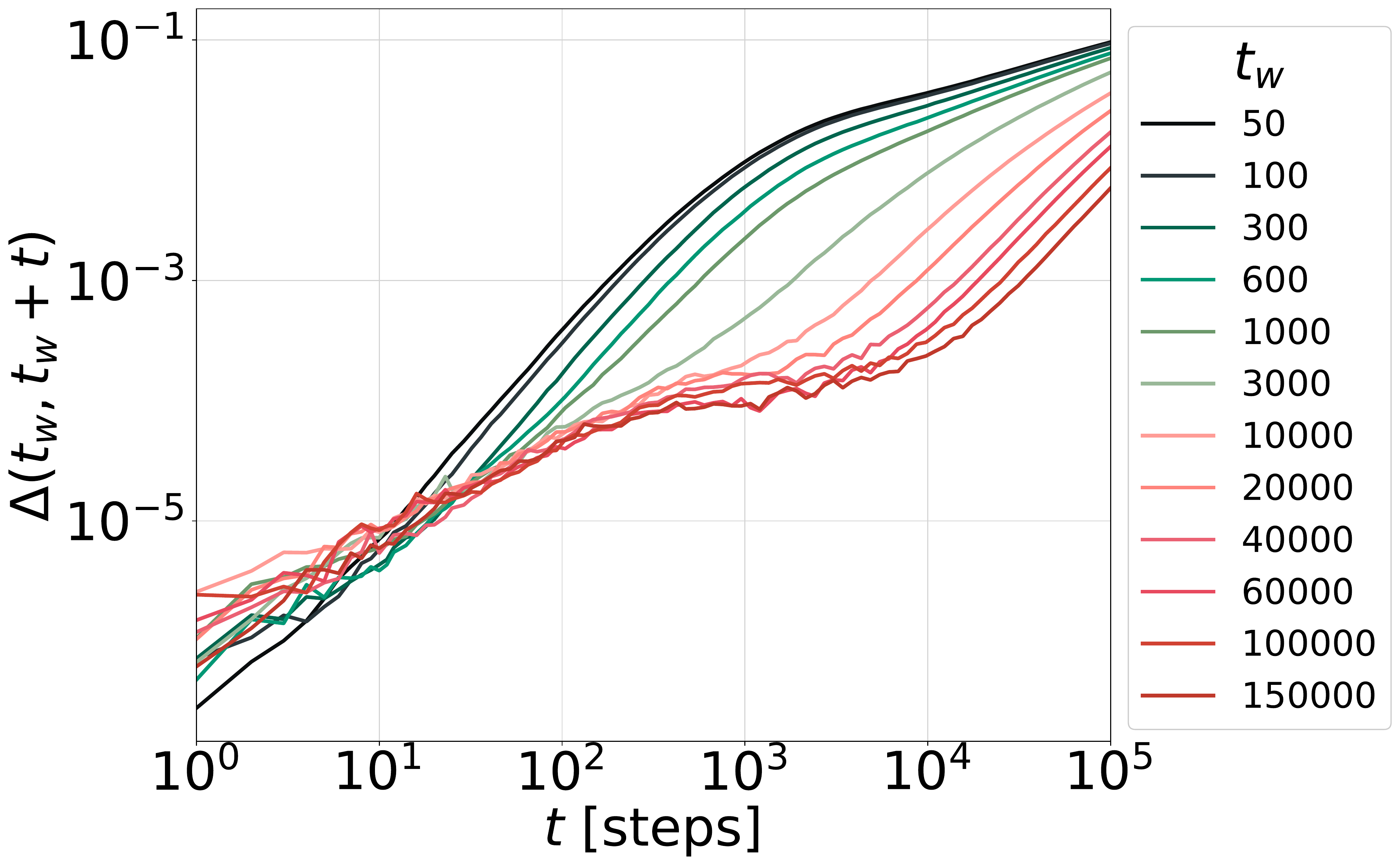}
         \caption{}
         \label{fig:corr2H}
     \end{subfigure}
     %\hfill
     
     \begin{subfigure}{0.47\textwidth}
         \centering
          \includegraphics[width=\textwidth]{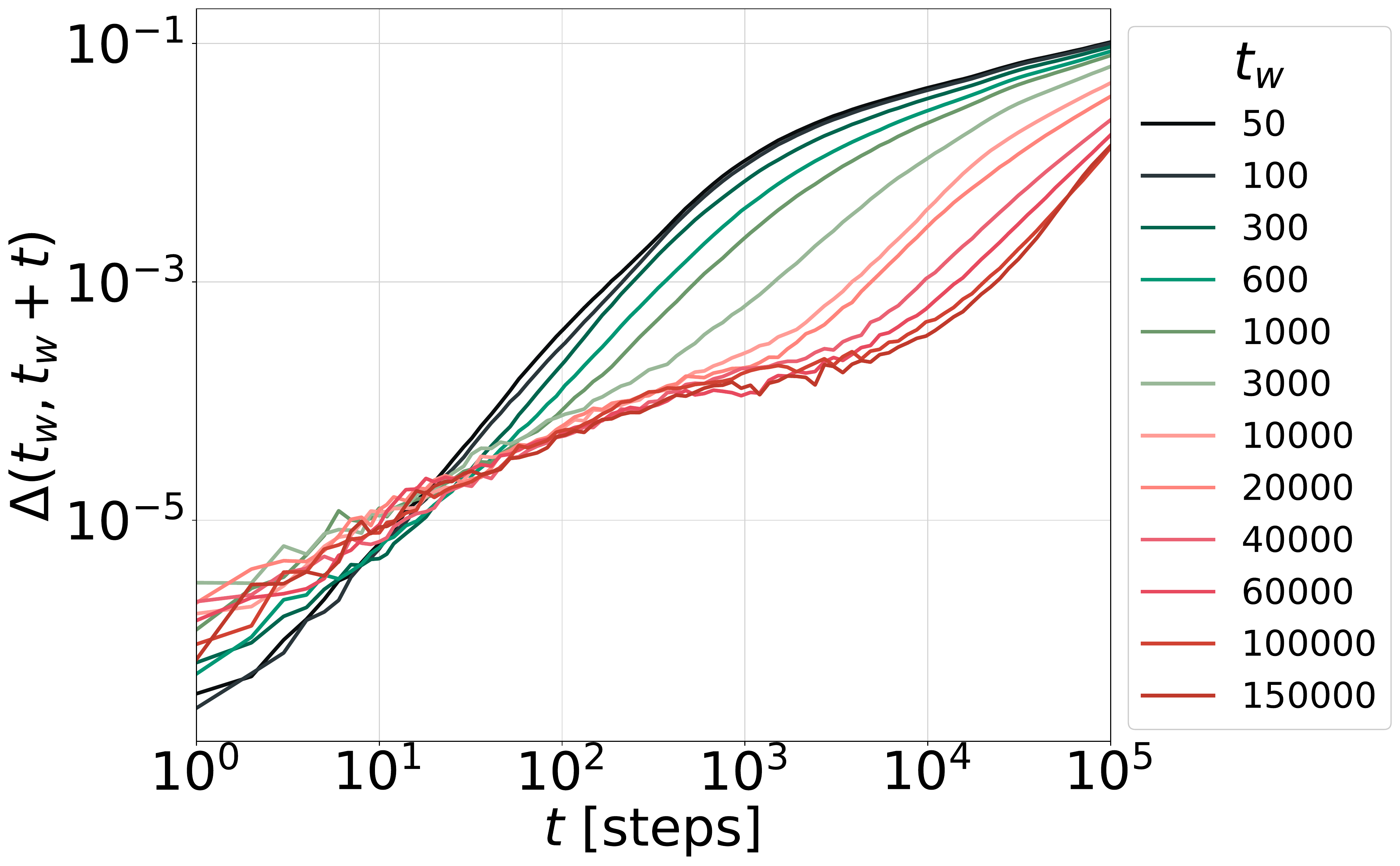}
         \caption{}
         \label{fig:corr3H}
     \end{subfigure}
     \hfill
     \begin{subfigure}{0.47\textwidth}
         \centering
         \includegraphics[width=\textwidth]{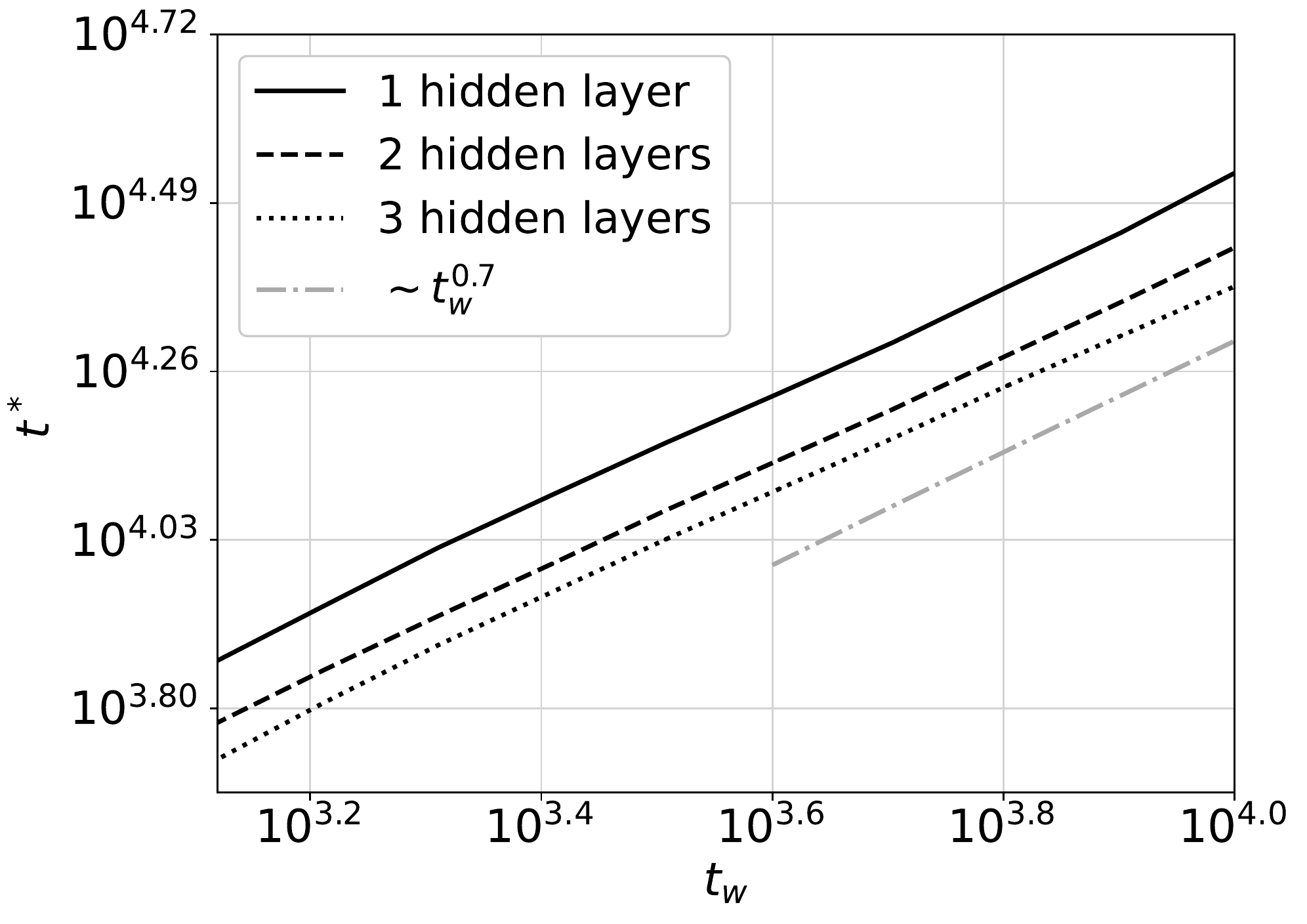}
         \caption{}
         \label{fig:twtmmist}
     \end{subfigure}

        \caption{Mean square displacement for MNIST dataset. Figure {\bf (a)} corresponds to the architecture with one hidden layer, {\bf (b)} to two hidden layers and {\bf (c)} to three hidden layers. Figure {\bf (d)} shows the characteristic time $t^*$ for each $t_{\rm{w}}$; for this plot, each curve is an average over 50 independent runs. The scaling with $t_{\rm{w}}$ shown as a dot-dashed line is set by hand with the purpose of illustration.}
        \label{fig:corrmnist}
\end{figure}

%With the results so far obtained, we conclude that the interplay between landscape and SGD gives rise to effects similar to those associated with structural glasses, namely a slow evolution of the loss (energy for a glass) and a $t_{\rm{w}}$ dependence of the mean square displacement. 

To show that our observations for the dynamics are quite generic in the underparametrized regime, we repeat the characterization for a minimal architecture (a chain) that contains three hidden layers and one neuron per layer. The landscape turns out to be rather simple, with a single pair of minima found with the basin--hopping algorithm~\footnote{We remark that the search is not exhaustive and new minima may appear with a longer run. We ran the GMNIN software for a time period of one day.}. Nevertheless, the training dynamics exhibit similar features to the higher--dimensional parameter space as shown in Figure~\ref{fig:lossmin}.

Additionally, we consider a more complex scenario and repeat the previous analysis using the MNIST database. We {fix the ratio $\alpha = P/N  \approx 0.07$ as investigated for the LJAT19 and use this as a constraint for different architectures with $H = 1,2, 3$ hidden layers.  As dataset we use the complete database, so that $N = 60000$. Therefore, $P \approx 4200$ (details in appendix~\ref{methods}).} Although for this scenario we do not have a characterization of the structure of the landscapes, we observe the same generic features in the evolution of the loss function, the MSD and the subaging behavior as for the LJAT19 database (see figures~\ref{fig:lossmnist} and~\ref{fig:corrmnist}). %\red{\st{Remarkably, the power--law exponent $\mu$ characterizing the subaging is very similar for the two considered databases. }}

\begin{figure}[t]
     \centering
     \begin{subfigure}[b]{0.55\textwidth}
         \centering
         \includegraphics[width=\textwidth]{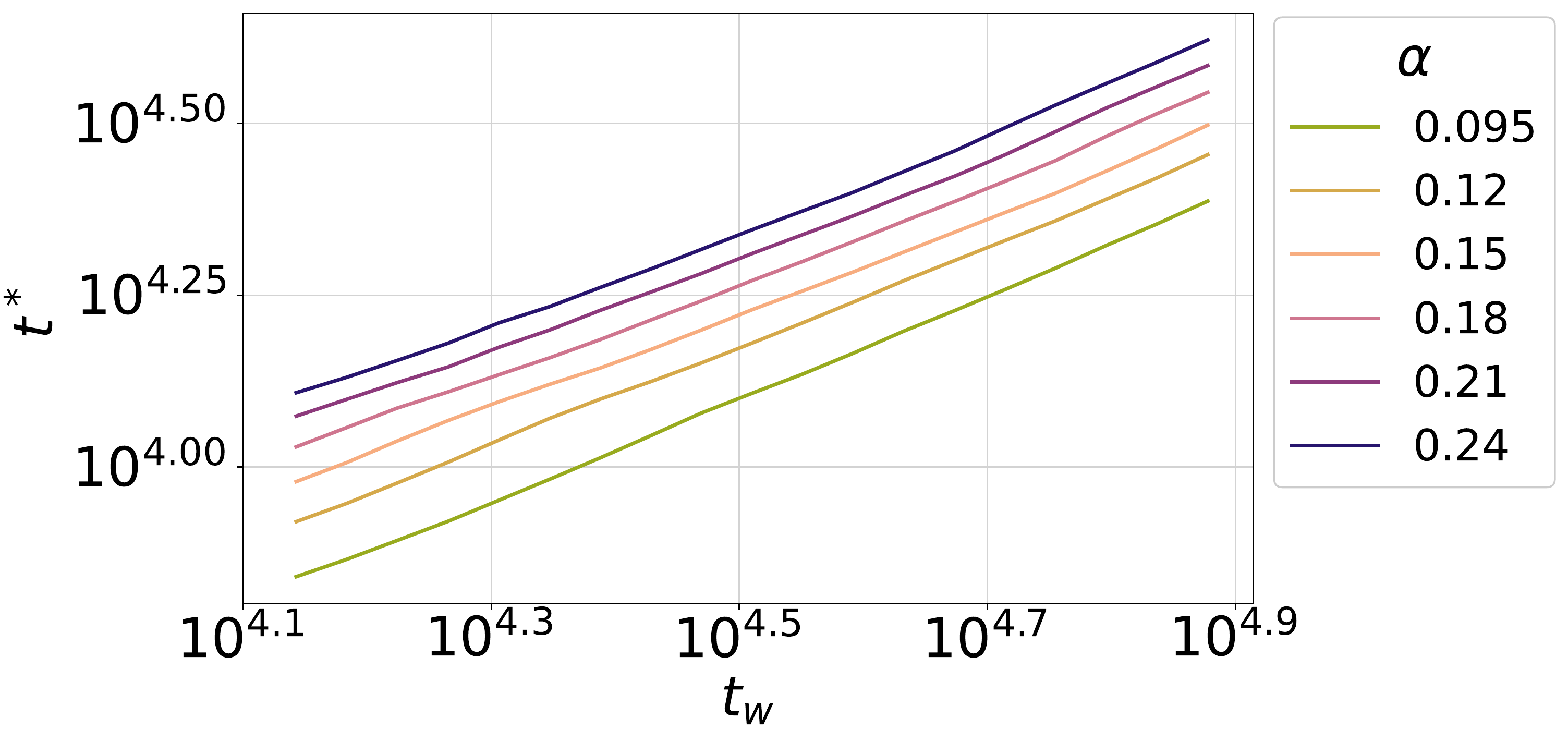}
         \caption{}
         \label{fig:tstartwmnist}
     \end{subfigure}
     \hfill
     \begin{subfigure}[b]{0.4\textwidth}
         \centering
         \includegraphics[width=\textwidth]{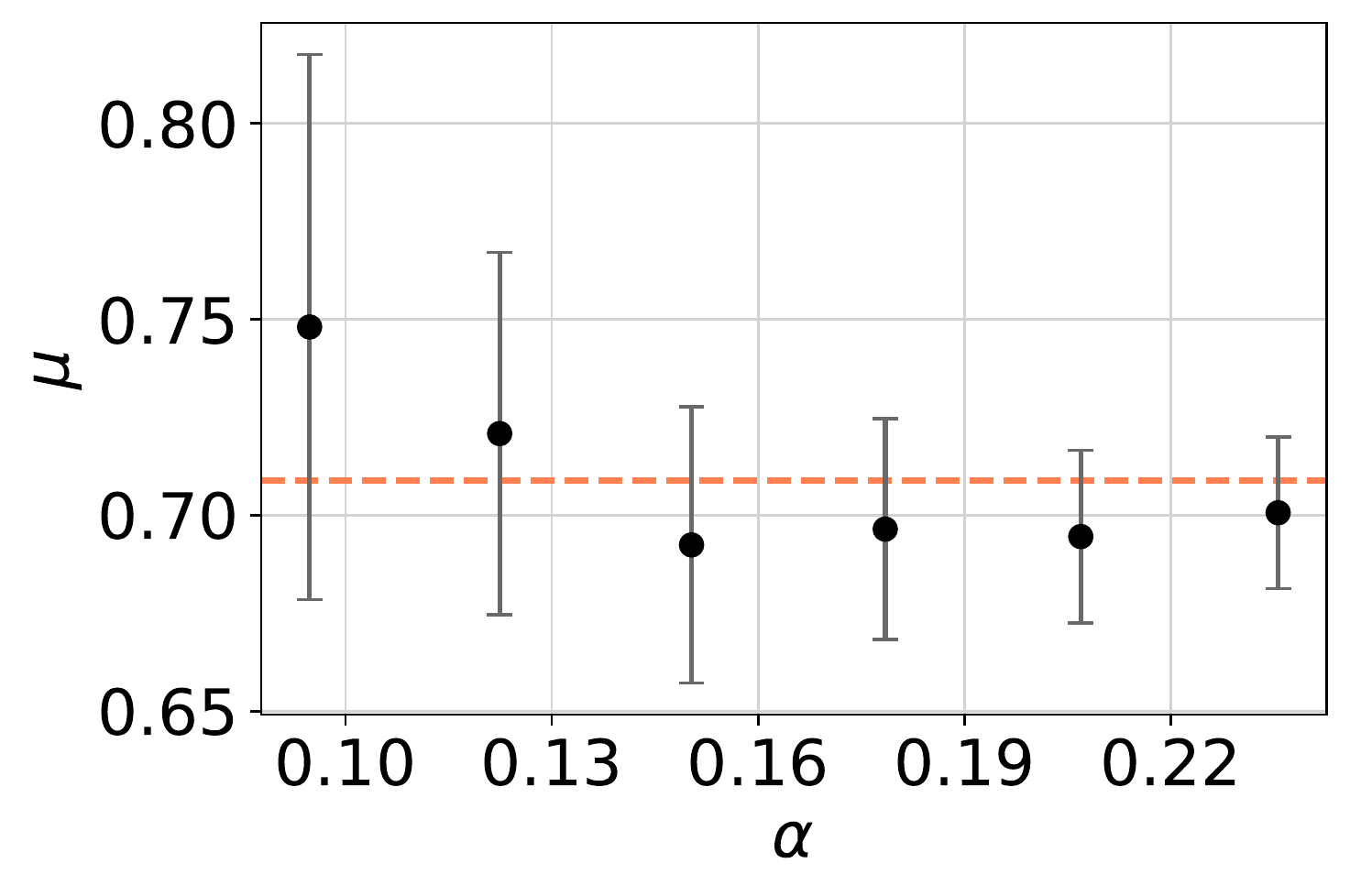}
         \caption{}
         \label{fig:exponentsmnist}
     \end{subfigure}
        \caption{{Characteristic time $t^*(t_{\rm{w}})$, extracted from the condition $\Delta^* = 10^{-3}$, for  a three hidden layers network and different values of $\alpha$ (increasing from bottom to top in (a)) for the MNIST database. Figure~{\bf (a)} corresponds to the mean $t^*$ over an ensemble of 100 independent networks for each $\alpha$. Figure~{\bf (b)} shows the corresponding exponent $t^* \sim t_{\rm{w}}^\mu$ extracted via least squares. The horizontal dashed line corresponds to the mean value,  $\mu = 0.709$.} }
        \label{fig:undermnist}
\end{figure}

{As a further step to characterize the underparametrized phase for the MNIST database, we study the scaling of the characteristic time with the waiting time for different values of $\alpha$ for a fixed depth $H = 3$.}~\footnote{{In practice, we consider that a network belongs to the underparametrized phase as long as $\alpha \ll 1$ and the accuracy reached after $t = 10^6$ steps is less than 0.99. The width of the networks is varied from $7$ to $17$ neurons, in steps of $2$.} } . {The data together with the associated exponent $\mu(\alpha)$ is shown in Figure~\ref{fig:undermnist}. We observe that the data is consistent with a unique exponent across the phase, that we found to be $\mu = 0.709 \pm 0.017$. }

\begin{figure}[b]
    \centering
     \begin{subfigure}{\textwidth}
         \centering
         \includegraphics[width=0.7\textwidth, scale=0.5]{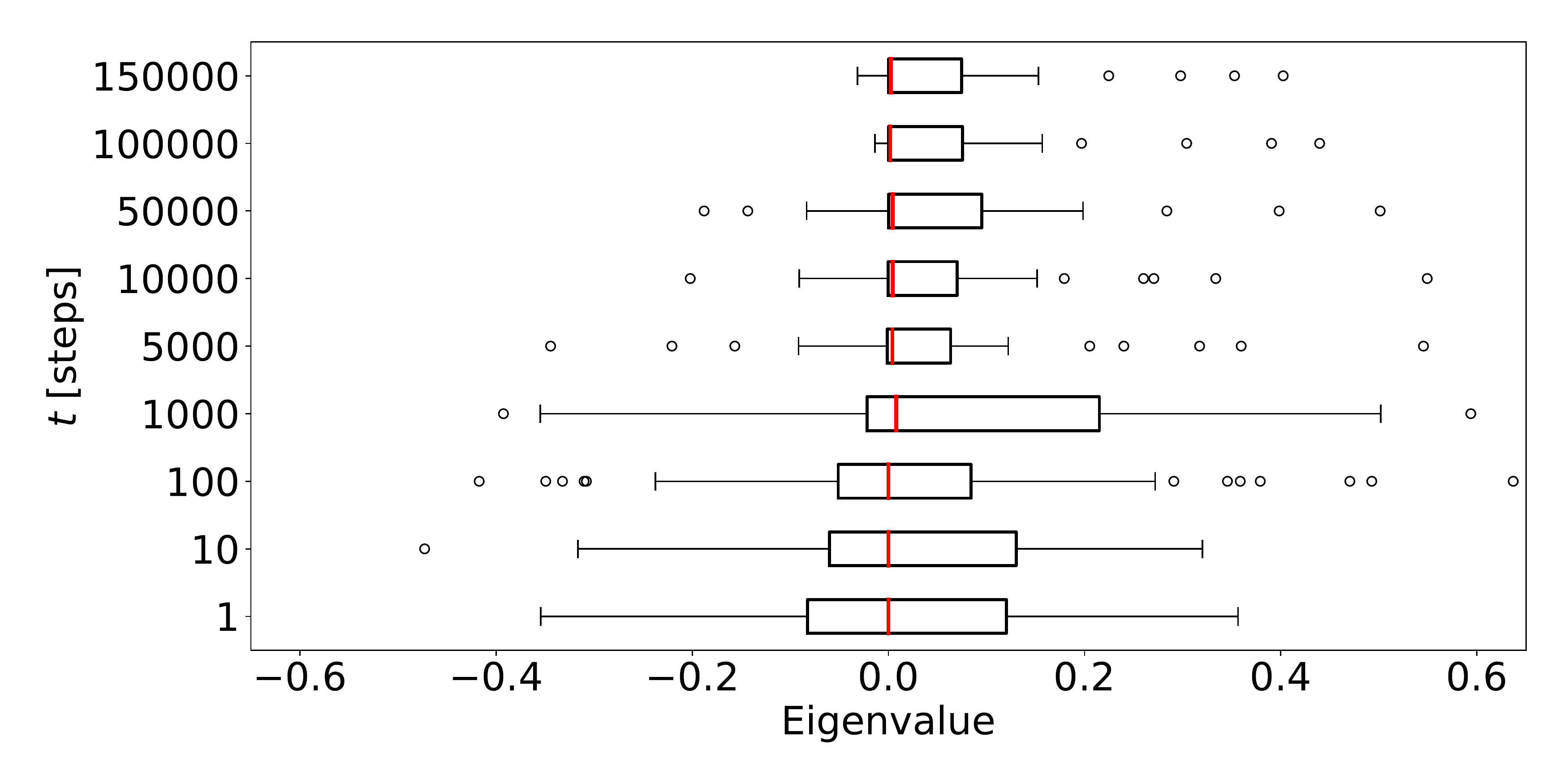}
         \caption{}
     \end{subfigure}
     
     \begin{subfigure}{\textwidth}
         \centering
         \includegraphics[width=0.7\textwidth, scale=0.5]{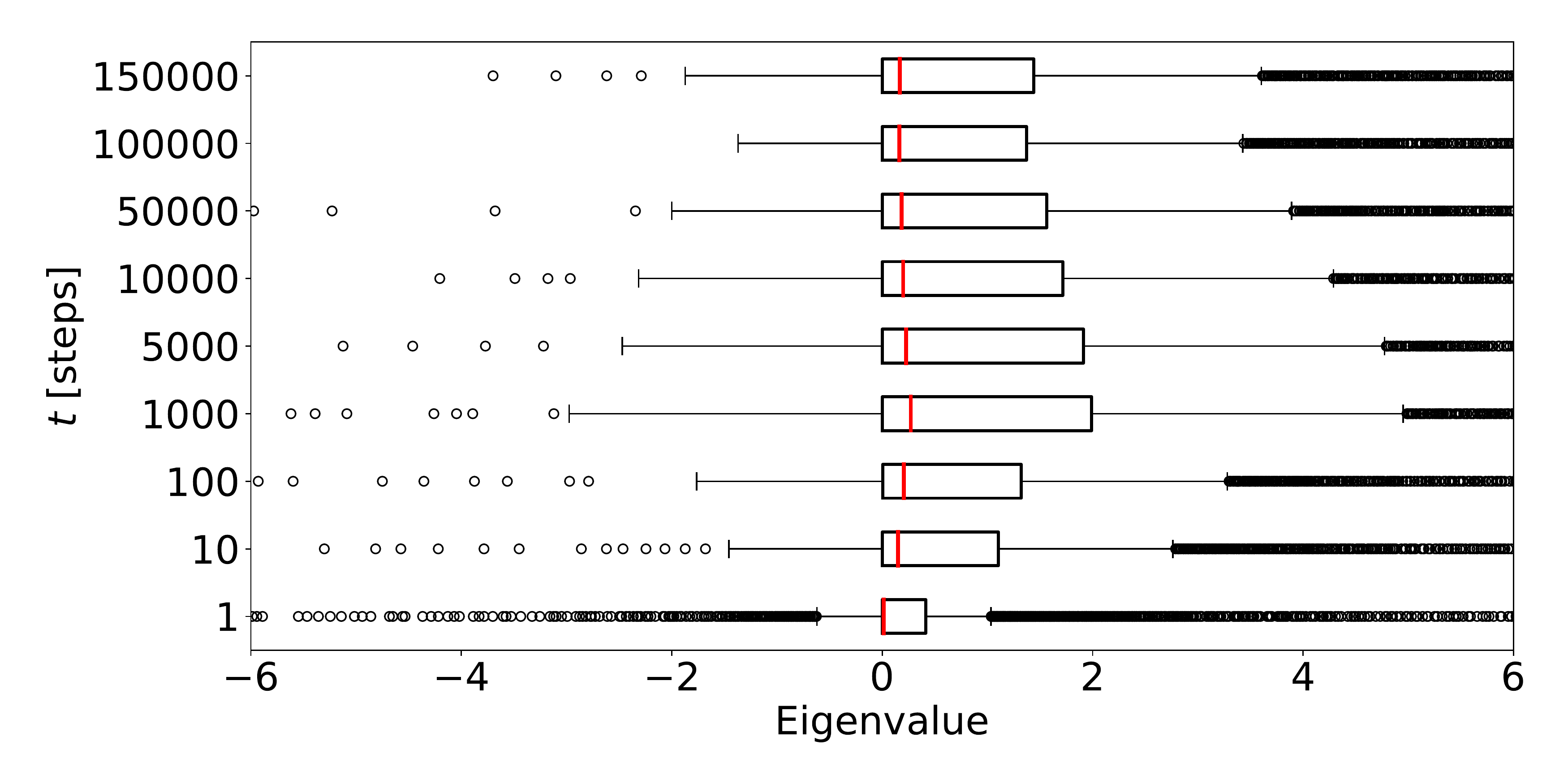}
         \caption{}
     \end{subfigure}
     \caption{Boxplots for the eigenvalues of the hessian at different training steps for the architecture with two hidden layers for the LJAT19 dataset (top) and the MNIST dataset (bottom). The boxes extend from the first to the third quartile, while the whiskers cover 1.5 times the interquartile range. For scalling, some outliers are not shown. It is evident there is a greater amount of eigenvalues that are close to zero or positive at the end of the training process.}
     \label{fig:boxplot}
\end{figure}

Finally, and in order to gain a better intuition on the interplay between landscape and dynamics, we explore the distribution of eigenvalues of the Hessian of the loss function during the training. This gives a notion of the local geometry of the landscape around the point where they are calculated~\cite{sagun2017empirical, wei2019noise}. In figure~\ref{fig:boxplot} we show boxplots of the distribution for different training steps for a single architecture and the two different databases. At the start of the training process, we see that the eigenvalues are distributed in a wide range of values. In contrast, in the end, the majority of the eigenvalues seem to be spread around zero, with some outliers being mainly positive. The number of negative eigenvalues decreases as training advances, but they do not disappear entirely. which is naturally a consequence of the stochasticity of the equations of motion (SGD). \footnote{We remark that we use a hyperbolic tangent as the activation function for the hidden layers (see methods in appendix~\ref{methods}). Therefore we do not expect cusps in the associated landscapes, and the negative eigenvalues can be attributed entirely to the dynamics.  This would not be the case with other activation functions, such as ReLU  (for a discussion we refer the reader to~\cite{geiger2019}). }

%%%%%%%%%%%%%%%%%%%%%%%%%%%%%%%%%%%%%%%%%%%%%%%%%%%%%%%%%%%%%%%%%%
\section{Conclusions}
\label{conc}

{Following the original observation in~\cite{Baity_Jesi_2019}, we show that the interplay between the structure of the landscape and the stochastic gradient descent in the underparametrized regime of a Deep Neural Network leads to glassy features. This glassiness is manifested  in the slow evolution (after a transient) of the loss function and the non time--translation invariant mean square displacement. In fact, for the cases considered we presented evidence of subaging, in the sense that the characteristic time grows more slowly than the age of the system.}

Furthermore, we show that the depth of the network plays a marginal role by comparing architectures with roughly the same number of parameters. In addition, along the lines of~\cite{fengPhases, frankle2020early}  we identify two phases of learning, and associate the dynamics of the last (slow) phase with the exploration of the bottom of the landscape where the deepest local minima lie.  Additionally, we state that the slow phase may be separated into two regimes according to the behavior of the MSD. The first regime exhibits subaging as characterized by a timescale that grows sublinearly with the waiting time. The second regime shows also  a watiting--time dependence of the MSD with a longer associated timescale. It is left for a future work to characterize the transition between regimes by computing, for instance, a generalized diffusion coefficient associated with each phase.

%\red{\st{It is also worth pinning down the dependence of the subaging exponent $\mu$ with the ratio $\alpha = P/N$   for a standard dataset, like MNIST. This could lead to a better understanding on the jamming transition from the underparametrized to the overparametrized regime}~\cite{geiger2019, geiger2021landscape}. }

%Along the lines set in references~\cite{geiger2019, spigler2019jamming, geiger2021landscape}, 
{As stated in references~\cite{geiger2019, spigler2019jamming, geiger2021landscape},  the transition from the underparametrized to the overparametrized regime is analogous to the solid--fluid change happening in amorphous materials with a variation of density (analogue of $1/\alpha$) and known as the \emph{jamming} transition~\cite{altieri2019jamming}.} {We characterize the underparametrized (jammed) phase for the MNIST, and our results point out to a unique subaging exponent across the whole phase. This is consistent with observations for amorphous solids obtained via athermal simulations of viscous soft particles (see for instance~\cite{head2009critical}). Our result is certainly intriguing and calls attention to a theoretical study that elaborates, firstly, on the origin of subaging in the underparametrized phase, secondly, on the nature of the scaling exponent, and thirdly, on the connection between the subaging behavior and other relevant properties of learning, such as the generalization error~\cite{spigler2019jamming}.}

{Finally, considering that} our results are robust to different landscapes, it is worth exploring the robustness to different types of noises. In this direction, an interesting option is to consider the \emph{persistent}--SGD~\cite{mignacco2021effective}, which has the appealing property of being amenable to a dynamical mean field theory treatment~\cite{mignacco2020dynamical}.

%Even though it is hard to say to what extent the observed properties depend on the landscape or dynamics, we have shown that the glassy properties are maintained in the simplistic scenario of only one neuron per hidden layer. To better characterise the role of the dynamics, we think it is worth considering alternative schemes as persistent--SGD~\cite{mignacco2020dynamical}.

\vspace{3mm} 

\emph{Acknowledgments.} We thank Peter Sollich, Moshir Harsh, Jack Parley, Rituparno Mandal and Marco Baity-Jesi for valuable discussions and feedback.  We also thank Philipp Verpoort and especially Max Niroomand for technical support.

\newpage

\section{Appendix}

\setcounter{figure}{0} \renewcommand{\thefigure}{A.\arabic{figure}}

\subsection{Methods}
\label{methods}

We worked with the architectures described in reference~\cite{Verpoort2020}, that is, three different configurations with one, two and three hidden layers, each with 10, 5 and 4 neurons per layer respectively. The total number of parameters is 74, 69 and 72 respectively. As mentioned in the main text, we also consider the same dataset LJAT19~\cite{ljat19data}. As input, we take two of the three coordinates that describe the triatomic system, and as output, one of the four possible stable fixed points (see the appendix of \cite{Verpoort2020} for more details). We used Tensorflow (v2.4) and the Keras API (v2.4) for training.

For the loss function, we used cross entropy with soft-max outputs and $\tanh{(x)}$ as activation for the hidden layers, as well as L2 regularization with a constant of $\lambda = 10^{-4}$. The initialization for each layer is Keras default initialization (Xavier uniform). We used SGD with a learning rate of 0.05, a batch size of 100 and the training is run through {150000} steps. 

For the minimum architecture, we put only one neuron in each hidden layer, resulting in 11, 13 and 15 parameters respectively. All other hyperparameters stay the same. For brevity, we show only the results for the case of three hidden layers.

For the MNIST dataset, 5 neurons are used in each hidden layer, resulting in architectures with  3985, 4015, 4045 parameters for one, two and three hidden layers respectively. The batch size is changed to 50 and all other hyperparameters are kept the same. The training is run through {300000} steps.

\begin{figure}[t]
    \centering
    \begin{subfigure}{0.43\textwidth}
         \centering
         \includegraphics[width=\textwidth, scale=0.5]{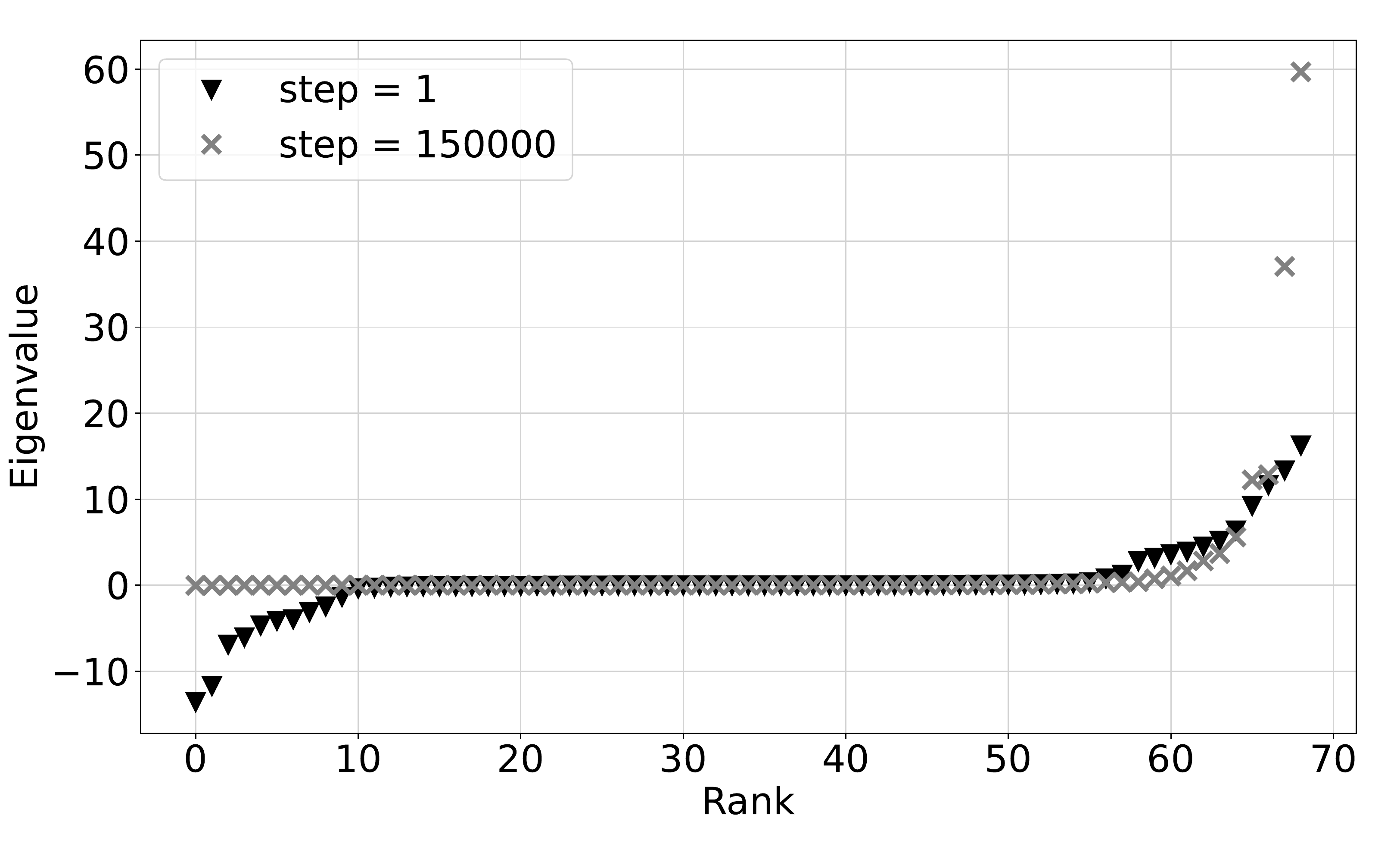}
         \caption{}
    \end{subfigure}
     \hfill
    \begin{subfigure}{0.45\textwidth}
         \centering
         \includegraphics[width=\textwidth, scale=0.5]{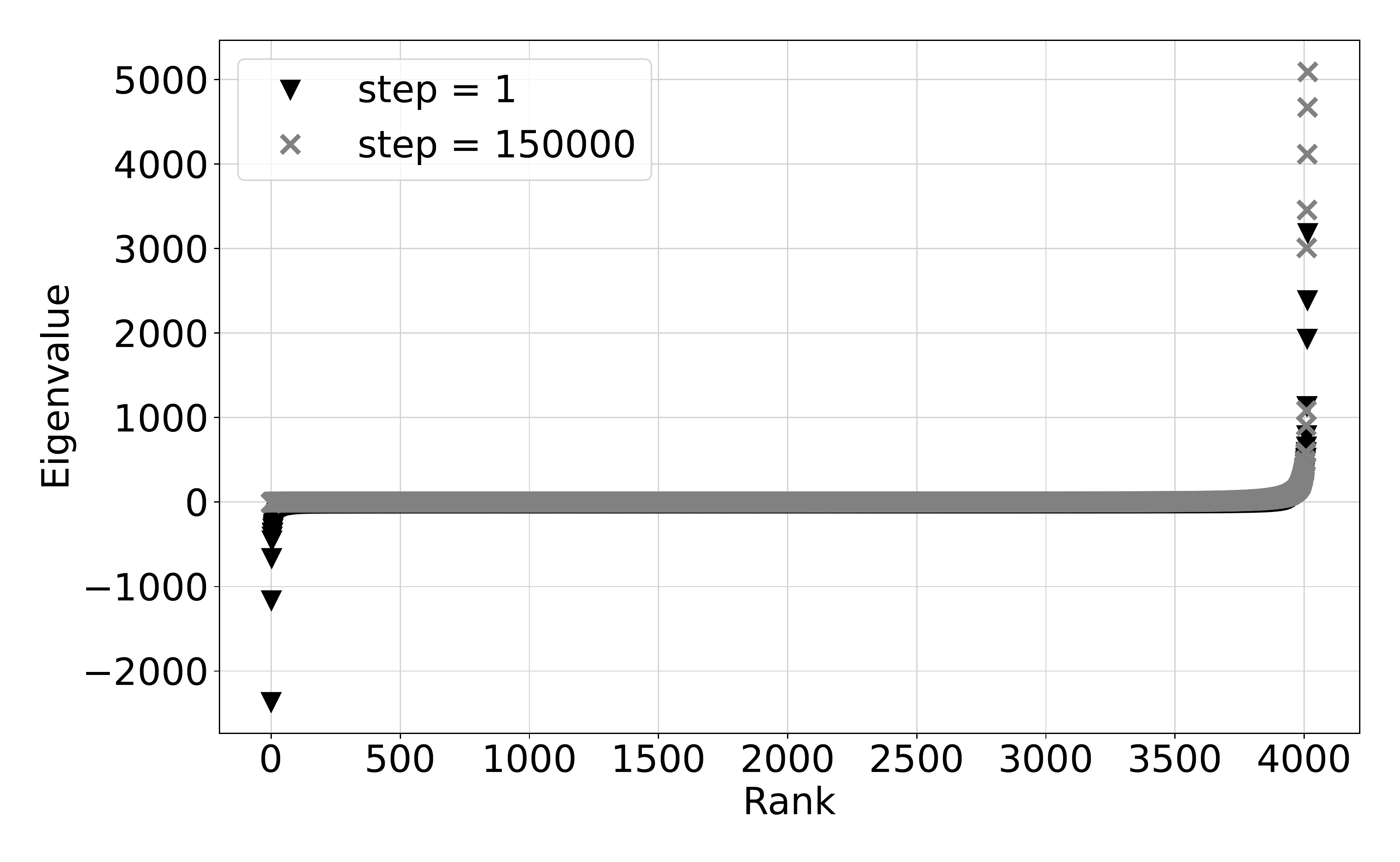}
         \caption{}
    \end{subfigure}
    \caption{Eigenvalues of the hessian at the start and end of training for the case of two hidden layers for the LJAT19 dataset (left) and the MNIST dataset (right). Here it is easier to visualize there is a decrease in the amount of negative eigenvalues at the end of the training process.}
    \label{fig:eigenvals}
\end{figure}

\subsection{Eigenvalues of the Hessian}

Following the results presented by \cite{sagun2017empirical}, we show the distribution of the hessian eigenvalues in the first and last training steps in Fig. \ref{fig:eigenvals}, ordered from least to greatest. This is a different way to visualize the change in distribution from the start to the end of the training process, where we can observe a reduction in the number of negative eigenvalues.

%Siguiendo a Sagun et al. mostramos la distribución de eigenvalues que reportamos en los boxplots de esta manera alternativa. Lo hacemos para ilustrar visualmente el cambio en la distribución. Tal vez mencionar el gap que separa el número de categorías del resto @@Agregar las figuras y hacer el caption@@

\begin{figure}[t]
     \centering
         \includegraphics[width=0.45\textwidth]{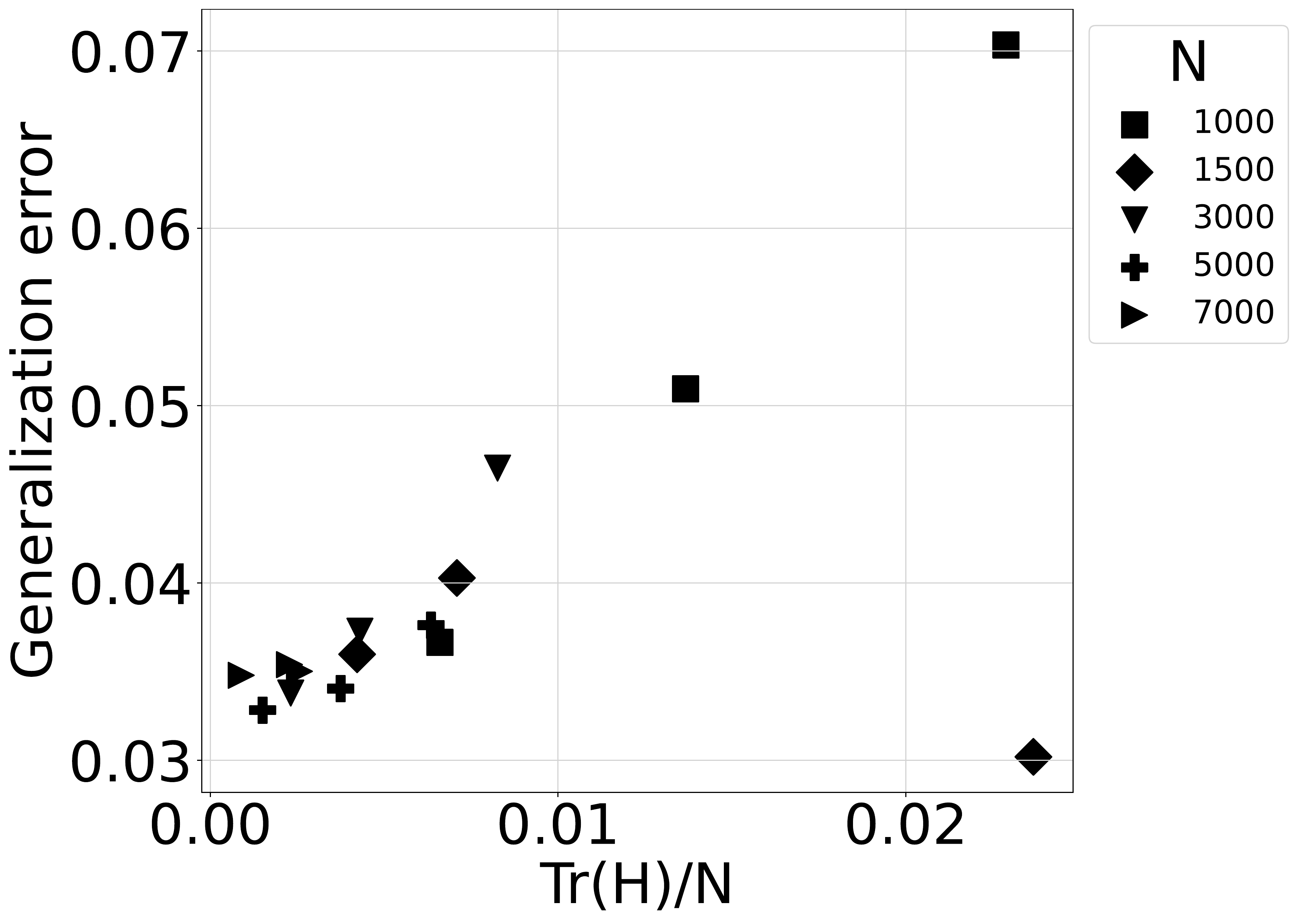}
         \label{fig:TrHN}
        \caption{{Trace of the hessian normalized by the dataset size for the architectures discussed in \cite{Verpoort2020}  against the generalization error at the end of the training, $t = {150000}$. Each marker indicates a different dataset size (described in the labels). There are three points for each marker type, corresponding to the architectures of one, two and three hidden layers.}}
        \label{fig:Thessian}
\end{figure}

{The trace of the hessian has been used as a way to estimate the generalization error of neural networks~\cite{MinimaSGD}. This error is defined as the difference between the test and training loss~\cite{bahri2020statistical}. With the intention of confirming the correlation found by \cite{Verpoort2020} between the local curvatures and the generalizability of minima, in figure \ref{fig:Thessian} we plot the generalization error against the normalized (by the dataset size) trace of the hessian for different dataset sizes at the last training step. The positive correlation supports the findings in the aforementioned reference.}

\subsection{Gradient}

\begin{figure}[t]
     \centering
         \includegraphics[width=0.5\textwidth]{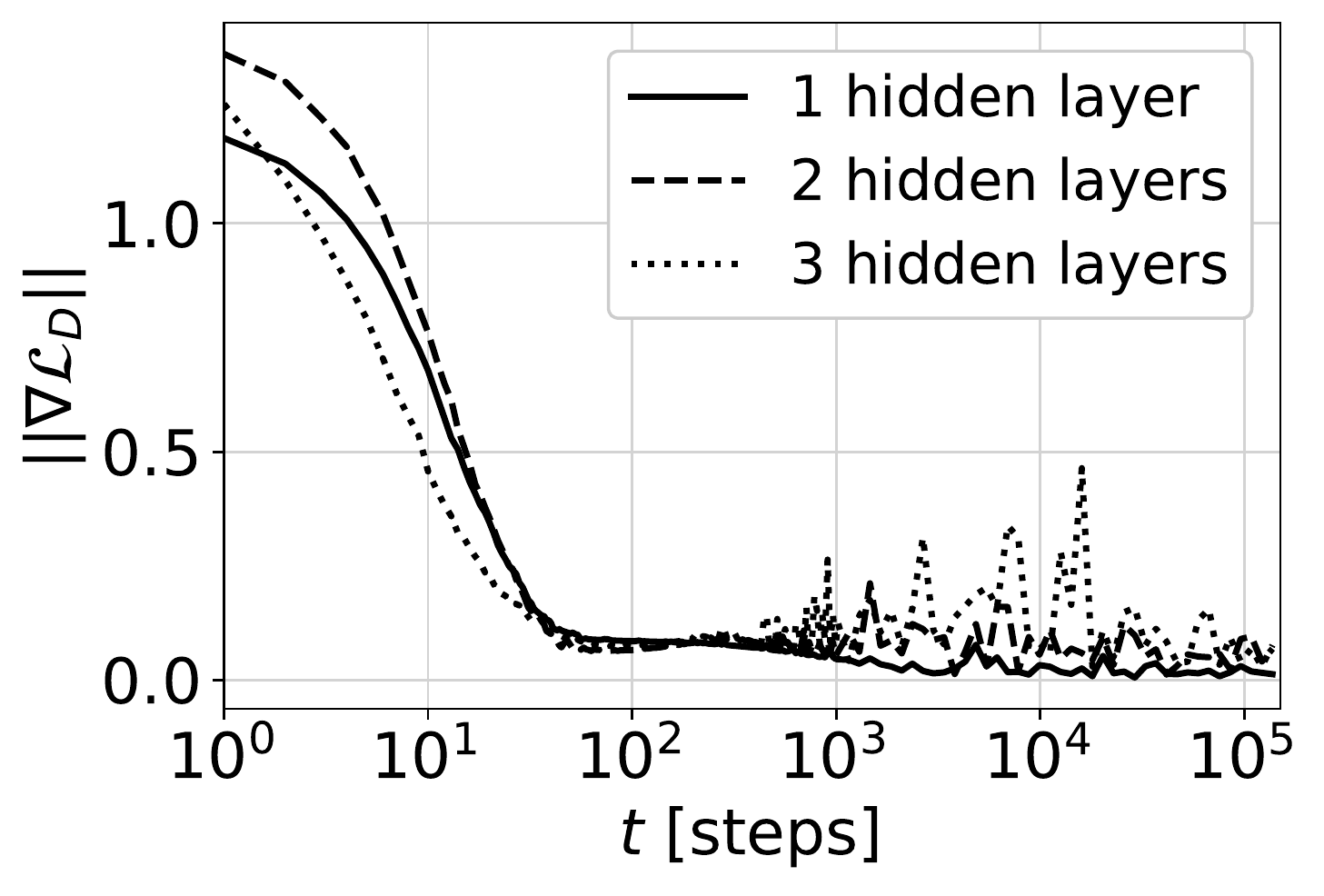}
         \label{fig:TrHN}
        \caption{{Evolution of the magnitude of the gradient of the loss for networks with one, two and three hidden layers associated with the database LJAT19 with $\alpha \approx 0.07$ and $N = 1000$.}}
        \label{fig:grads}
\end{figure}

{Using the same architectures discussed by \cite{Verpoort2020}, we calculate the magnitude of the gradient of the loss function in each training step (figure \ref{fig:grads}). In general, we observe a fast decrease to a value close to zero, around which it fluctuates. The noise in this part is dependent on the depth of the network, being more prominent for the network with three hidden layers and less so for the one with one hidden layer. The root-mean-square of the points after the step $t = 10^4$ is 0.02, 0.07 and 0.12 for one, two and three hidden layers respectively. This result can be rationalized from the perspective of the roughness of the landscape. As the network becomes deeper, the number of (shallow) local minima increases, and hence the stochastic dynamics induces higher fluctuations of the gradient during the last phase of the training.}

\newpage

\bibliographystyle{apalike}
\bibliography{bibliography}

\begin{thebibliography}{}

\bibitem[Altieri, 2019]{altieri2019jamming}
Altieri, A. (2019).
\newblock The jamming transition.
\newblock In {\em Jamming and Glass Transitions}, pages 45--64. Springer.

\bibitem[Arceri et~al., 2020]{arceri2020glasses}
Arceri, F., Landes, F.~P., Berthier, L., and Biroli, G. (2020).
\newblock Glasses and aging: a statistical mechanics perspective.
\newblock {\em arXiv preprint arXiv:2006.09725}.

\bibitem[Bahri et~al., 2020]{bahri2020statistical}
Bahri, Y., Kadmon, J., Pennington, J., Schoenholz, S.~S., Sohl-Dickstein, J.,
  and Ganguli, S. (2020).
\newblock Statistical mechanics of deep learning.
\newblock {\em Annual Review of Condensed Matter Physics}, 11(1).

\bibitem[Baity-Jesi et~al., 2019]{Baity_Jesi_2019}
Baity-Jesi, M., Sagun, L., Geiger, M., Spigler, S., Arous, G.~B., Cammarota,
  C., LeCun, Y., Wyart, M., and Biroli, G. (2019).
\newblock Comparing dynamics: deep neural networks versus glassy systems.
\newblock {\em Journal of Statistical Mechanics: Theory and Experiment},
  2019(12):124013.

\bibitem[Ballard et~al., 2016]{ballard2016energy}
Ballard, A.~J., Stevenson, J.~D., Das, R., and Wales, D.~J. (2016).
\newblock Energy landscapes for a machine learning application to series data.
\newblock {\em The Journal of chemical physics}, 144(12):124119.

\bibitem[Berthier, 2000]{berthier2000sub}
Berthier, L. (2000).
\newblock Sub-aging in a domain growth model.
\newblock {\em The European Physical Journal B-Condensed Matter and Complex
  Systems}, 17(4):689--692.

\bibitem[Bouchaud, 1999]{bouchaud1999aging}
Bouchaud, J.-P. (1999).
\newblock Aging in glassy systems: new experiments, simple models, and open
  questions.
\newblock {\em arXiv preprint cond-mat/9910387}.

\bibitem[Carleo et~al., 2019]{carleo2019machine}
Carleo, G., Cirac, I., Cranmer, K., Daudet, L., Schuld, M., Tishby, N.,
  Vogt-Maranto, L., and Zdeborov{\'a}, L. (2019).
\newblock Machine learning and the physical sciences.
\newblock {\em Reviews of Modern Physics}, 91(4):045002.

\bibitem[Chen et~al., 2022]{chen2022anomalous}
Chen, G., Qu, C.~K., and Gong, P. (2022).
\newblock Anomalous diffusion dynamics of learning in deep neural networks.
\newblock {\em Neural Networks}, 149:18--28.

\bibitem[Choromanska et~al., 2015]{choromanska2015loss}
Choromanska, A., Henaff, M., Mathieu, M., Arous, G.~B., and LeCun, Y. (2015).
\newblock The loss surfaces of multilayer networks.
\newblock In {\em Artificial intelligence and statistics}, pages 192--204.
  PMLR.

\bibitem[de~Souza and Wales, 2008]{de2008energy}
de~Souza, V.~K. and Wales, D.~J. (2008).
\newblock Energy landscapes for diffusion: Analysis of cage-breaking processes.
\newblock {\em The Journal of chemical physics}, 129(16):164507.

\bibitem[Feng and Tu, 2021]{fengPhases}
Feng, Y. and Tu, Y. (2021).
\newblock Phases of learning dynamics in artificial neural networks: with or
  without mislabeled data.

\bibitem[Fielding et~al., 2000]{fielding2000aging}
Fielding, S.~M., Sollich, P., and Cates, M.~E. (2000).
\newblock Aging and rheology in soft materials.
\newblock {\em Journal of Rheology}, 44(2):323--369.

\bibitem[Frankle et~al., 2020]{frankle2020early}
Frankle, J., Schwab, D.~J., and Morcos, A.~S. (2020).
\newblock The early phase of neural network training.
\newblock {\em arXiv preprint arXiv:2002.10365}.

\bibitem[Franz and Parisi, 2016]{franz2016simplest}
Franz, S. and Parisi, G. (2016).
\newblock The simplest model of jamming.
\newblock {\em Journal of Physics A: Mathematical and Theoretical},
  49(14):145001.

\bibitem[Geiger et~al., 2021]{geiger2021landscape}
Geiger, M., Petrini, L., and Wyart, M. (2021).
\newblock Landscape and training regimes in deep learning.
\newblock {\em Physics Reports}, 924:1--18.

\bibitem[Geiger et~al., 2019]{geiger2019}
Geiger, M., Spigler, S., d'Ascoli, S., Sagun, L., Baity-Jesi, M., Biroli, G.,
  and Wyart, M. (2019).
\newblock Jamming transition as a paradigm to understand the loss landscape of
  deep neural networks.
\newblock {\em Phys. Rev. E}, 100:012115.

\bibitem[Head, 2009]{head2009critical}
Head, D.~A. (2009).
\newblock Critical scaling and aging in cooling systems near the jamming
  transition.
\newblock {\em Physical review letters}, 102(13):138001.

\bibitem[Jacot et~al., 2018]{jacot2018neural}
Jacot, A., Gabriel, F., and Hongler, C. (2018).
\newblock Neural tangent kernel: Convergence and generalization in neural
  networks.
\newblock {\em Advances in neural information processing systems}, 31.

\bibitem[Jastrzębski et~al., 2017]{MinimaSGD}
Jastrzębski, S., Kenton, Z., Arpit, D., Ballas, N., Fischer, A., Bengio, Y.,
  and Storkey, A. (2017).
\newblock Three factors influencing minima in sgd.

\bibitem[Kunin et~al., 2021]{kunin2021limiting}
Kunin, D., Sagastuy-Brena, J., Gillespie, L., Margalit, E., Tanaka, H.,
  Ganguli, S., and Yamins, D.~L. (2021).
\newblock Limiting dynamics of sgd: Modified loss, phase space oscillations,
  and anomalous diffusion.
\newblock {\em arXiv preprint arXiv:2107.09133}.

\bibitem[LeCun, 1998]{lecun1998mnist}
LeCun, Y. (1998).
\newblock The mnist database of handwritten digits.
\newblock {\em http://yann. lecun. com/exdb/mnist/}.

\bibitem[Mehta et~al., 2019]{mehta2019high}
Mehta, P., Bukov, M., Wang, C.-H., Day, A.~G., Richardson, C., Fisher, C.~K.,
  and Schwab, D.~J. (2019).
\newblock A high-bias, low-variance introduction to machine learning for
  physicists.
\newblock {\em Physics reports}, 810:1--124.

\bibitem[Mei et~al., 2018]{mei2018mean}
Mei, S., Montanari, A., and Nguyen, P.-M. (2018).
\newblock A mean field view of the landscape of two-layer neural networks.
\newblock {\em Proceedings of the National Academy of Sciences},
  115(33):E7665--E7671.

\bibitem[Mignacco et~al., 2020]{mignacco2020dynamical}
Mignacco, F., Krzakala, F., Urbani, P., and Zdeborov{\'a}, L. (2020).
\newblock Dynamical mean-field theory for stochastic gradient descent in
  gaussian mixture classification.
\newblock {\em Advances in Neural Information Processing Systems},
  33:9540--9550.

\bibitem[Mignacco and Urbani, 2021]{mignacco2021effective}
Mignacco, F. and Urbani, P. (2021).
\newblock The effective noise of stochastic gradient descent.
\newblock {\em arXiv preprint arXiv:2112.10852}.

\bibitem[Niblett et~al., 2017]{niblett2017pathways}
Niblett, S., Biedermann, M., Wales, D., and De~Souza, V. (2017).
\newblock Pathways for diffusion in the potential energy landscape of the
  network glass former sio2.
\newblock {\em The Journal of Chemical Physics}, 147(15):152726.

\bibitem[Niblett et~al., 2016]{niblett2016dynamics}
Niblett, S., De~Souza, V., Stevenson, J., and Wales, D. (2016).
\newblock Dynamics of a molecular glass former: {E}nergy landscapes for
  diffusion in ortho-terphenyl.
\newblock {\em The Journal of chemical physics}, 145(2):024505.

\bibitem[Niroomand, 2021]{maxpylfl}
Niroomand, M. (2021).
\newblock pylfl: A tool for loss function landscape exploration in python.
\newblock \url{https://github.com/orinxam/pylfl}.

\bibitem[Rinn et~al., 2000]{rinn2000multiple}
Rinn, B., Maass, P., and Bouchaud, J.-P. (2000).
\newblock Multiple scaling regimes in simple aging models.
\newblock {\em Physical review letters}, 84(23):5403.

\bibitem[Sagun et~al., 2017]{sagun2017empirical}
Sagun, L., Evci, U., Guney, V.~U., Dauphin, Y., and Bottou, L. (2017).
\newblock Empirical analysis of the hessian of over-parametrized neural
  networks.
\newblock {\em arXiv preprint arXiv:1706.04454}.

\bibitem[Scalliet and Berthier, 2019]{scalliet2019rejuvenation}
Scalliet, C. and Berthier, L. (2019).
\newblock Rejuvenation and memory effects in a structural glass.
\newblock {\em Physical review letters}, 122(25):255502.

\bibitem[Seung et~al., 1992]{seung1992statistical}
Seung, H.~S., Sompolinsky, H., and Tishby, N. (1992).
\newblock Statistical mechanics of learning from examples.
\newblock {\em Physical review A}, 45(8):6056.

\bibitem[Spigler et~al., 2019]{spigler2019jamming}
Spigler, S., Geiger, M., d’Ascoli, S., Sagun, L., Biroli, G., and Wyart, M.
  (2019).
\newblock A jamming transition from under-to over-parametrization affects
  generalization in deep learning.
\newblock {\em Journal of Physics A: Mathematical and Theoretical},
  52(47):474001.

\bibitem[Verpoort et~al., 2020a]{ljat19data}
Verpoort, P., Lee, A., and Wales, D. (2020a).
\newblock {LJAT19 dataset}.
\newblock \url{https://www.repository.cam.ac.uk/handle/1810/308755}.

\bibitem[Verpoort et~al., 2020b]{Verpoort2020}
Verpoort, P.~C., Lee, A.~A., and Wales, D.~J. (2020b).
\newblock Archetypal landscapes for deep neural networks.
\newblock {\em Proceedings of the National Academy of Sciences},
  117(36):21857--21864.

\bibitem[Wales et~al., 1999]{wales1999gmin}
Wales, D. et~al. (1999).
\newblock Gmin: A program for basin-hopping global optimisation,
  basin-sampling, and parallel tempering.
\newblock {\em See http://www-wales. ch. cam. ac. uk/software. html}.

\bibitem[Wales and Bogdan, 2006]{wales2006potential}
Wales, D.~J. and Bogdan, T.~V. (2006).
\newblock Potential energy and free energy landscapes.

\bibitem[Wales et~al., 1998]{wales1998archetypal}
Wales, D.~J., Miller, M.~A., and Walsh, T.~R. (1998).
\newblock Archetypal energy landscapes.
\newblock {\em Nature}, 394(6695):758--760.

\bibitem[Wei and Schwab, 2019]{wei2019noise}
Wei, M. and Schwab, D.~J. (2019).
\newblock How noise affects the hessian spectrum in overparameterized neural
  networks.
\newblock {\em arXiv preprint arXiv:1910.00195}.

\end{thebibliography}

\end{document}